\documentclass[aps,twocolumn,showpacs,showkeys,nofootinbib,superscriptaddress]{revtex4-1}

\usepackage{epsfig}
\usepackage{amsmath}
\usepackage{amsfonts}
\usepackage{amssymb}
\usepackage{mathtools}
\usepackage{graphicx}
\usepackage{colordvi}
\usepackage[dvipsnames]{xcolor}
\usepackage{xspace}
\usepackage{acronym}
\usepackage{booktabs}
\usepackage{hyperref}
\hypersetup{
  colorlinks=true,        
  linkcolor=blue,         
  citecolor=cyan,         
}
\usepackage{natbib}
\usepackage[final]{showlabels}

\graphicspath{{./}}

\newcommand{\dd}[0]{\textnormal{d}}
\newcommand{\bfs}[1]{\boldsymbol{#1}}
\newcommand{\ext}{\bfs{d}}
\newcommand{\covext}{\bfs{D}}
\newcommand{\base}{\bfs{\theta}}
\newcommand{\cform}{\bfs{\omega}}
\newcommand{\half}{\frac{1}{2}}
\newcommand{\sgn}{\text{sgn}}

\newcommand{\secref}[1]{Sec.~\ref{sec:#1}}

\newcommand{\dgrem}{\lowercase{d}\textsc{GREM}\@\xspace}
\newcommand{\adm}{\textsc{ADM}\@\xspace}
\newcommand{\bssnok}{\textsc{BSSNOK}\@\xspace}
\newcommand{\zfourc}{\textsc{Z4}\lowercase{c}\@\xspace}
\newcommand{\cczfour}{\textsc{CCZ4}\@\xspace}
\newcommand{\focczfour}{\textsc{FO-CCZ4}\@\xspace}

\def\ie{i.e.\@\xspace}
\def\eg{e.g.\@\xspace}
\def\cf{cf.\@\xspace}

\newacro{EM}{electrodynamics}
\newacro{MHD}{magnetohydrodynamics}
\newacro{BBH}{binary black hole}
\newacro{BNS}{binary neutron star}
\newacro{BH}{black hole}
\newacro{GW}{gravitational wave}
\newacro{GRB}{gamma ray burst}
\newacro{NR}{numerical relativity}
\newacro{EFE}{Einstein's field equations}
\newacro{IBVP}{initial boundary value problem}
\newacro{PIC}{particle-in-cell}
\newacro{CT}{constrained transport}
\newacro{dG}{discontinuous Galerkin}
\newacro{AMR}{adaptive mesh refinement}
\newacro{GR}{general relativity}
\newacro{PDE}{partial differential equation}
\newacro{GHG}{generalized-harmonic}

\begin{document}

\title{A new first-order formulation of the Einstein
equations exploiting analogies with electrodynamics}

\author{H. Olivares}
\email{h.olivares@astro.ru.nl}
\affiliation{Department of Astrophysics/IMAPP, Radboud University Nijmegen,
	P.O. Box 9010, NL-6500 GL Nijmegen, The Netherlands}

\author{I. M. Peshkov}
\affiliation{Department of Civil, Environmental and Mechanical Engineering, University
  of Trento, via Mesiano 77, 38123 Trento, Italy}

\author{E. R. Most}
\affiliation{Princeton Center for Theoretical Science, Princeton University, Princeton, NJ 08544, USA}
\affiliation{Princeton Gravity Initiative, Princeton University, Princeton, NJ 08544, USA}
\affiliation{School of Natural Sciences, Institute for Advanced Study, Princeton, NJ 08540, USA}

\author{F. M. Guercilena}
\noaffiliation

\author{L. J. Papenfort}
\affiliation{Institut für Theoretische Physik, Goethe Universität, Max-von-Laue-Str. 1,
  60438 Frankfurt am Main, Germany
}
\date{\today}

\begin{abstract}
When formulated as an initial boundary value problem, the Einstein and Maxwell
equations are both systems of hyperbolic equations for which variables need to
satisfy a set of elliptic constraints throughout evolution. However, while
\ac{EM} and \ac{MHD} have benefited from a large number of evolution schemes
that are able to enforce these constraints and are easily applicable to
curvilinear coordinates, unstructured meshes, or $N$-body (or particle-in-cell)
simulations, many of these techniques cannot be straightforwardly applied to
existing formulations of the Einstein equations. With the aim of building a
numerical scheme that exploits this existing technology, we develop a 3+1 a
formulation of the Einstein equations which shows a striking formal resemblance
to the equations of relativistic \ac{MHD} and to \ac{EM} in material media. The
fundamental variables of this formulation are the frame fields, their exterior
derivatives, and the Nester-Witten and Sparling forms. These mirror the roles of
the electromagnetic 4-potential, the electromagnetic field strengths, the field
excitations and the electric (in this case energy-momentum) current,
respectively. It also possess the lapse function and shift vector as gauge
freedoms, whose role corresponds exactly to that of the scalar part of the
electromagnetic 4-potential. The formulation, that we name \dgrem (for
\underline{d}ifferential forms, \underline{g}eneral \underline{r}elativity and
\underline{e}lectro\underline{m}agnetism), is manifestly first order and
flux-conservative, which makes it suitable for high-resolution shock capturing
schemes and finite-element methods. Being derived using techniques from exterior
calculus, it does not contain covariant but only exterior derivatives, which
makes it directly applicable to any coordinate system and to unstructured
meshes, and leads to a natural discretization in staggered grids potentially
suitable for the use of well-known techniques for constraint preservation such
as the Yee algorithm and constrained transport. Due to these properties, we
expect this new formulation to be beneficial in simulations of many astrophysical
systems, such as binary compact objects and core-collapse supernovae as well as
cosmological simulations of the early universe. However we leave its numerical
implementation for future work.
\end{abstract}

\maketitle

\section{Introduction}

In the last few years the study of relativistic astrophysics and in particular
of compact objects has made significant progress. The theoretical understanding
of \acp{BBH}, \acp{BNS} and super-massive \aclp{BH} has been validated by a
string of impressive observations, such as the first detection of \acp{GW} from
\acp{BBH}\citep{2016PhRvL.116f1102A}; the first and joint detection of \acp{GW},
a \ac{GRB} and a kilonova from a \ac{BNS} system \citep{2017PhRvL.119p1101A,
  2017ApJ...848L..12A}; and the first direct imaging of a super-massive
accreting \ac{BH}\citep{2019ApJ...875L...4E, 2021ApJ...910L..12E}.

These are systems exhibiting extreme complexity, and whose modeling requires the
interplay of different areas of modern physics, such as relativistic
gravitation, fluid dynamics, electrodynamics, nuclear physics, neutrino physics
and many others. Therefore the theoretical study of these and other systems
cannot be accomplished with purely analytical tools. \Ac{NR} has instead emerged
as a powerful modeling tool.

The core approach of \ac{NR} consists in finding approximate solutions to the
\acp{PDE} describing the system at study, namely the \ac{EFE}, by numerical
integration. To this end, the equations of \ac{GR} have first to be recast as an
\ac{IBVP}. This can be accomplished in various ways. Examples include the
generalized-harmonic formalism \citep{2005PhRvL..95l1101P, 2002PhRvD..65d4029G,
  2006CQGra..23S.447L}; the characteristic-evolution formalism
\citep{2012LRR....15....2W}; the conformal approach \citep{Frauendiener:2002,
  Husa:2003} and fully-constrained formulations \citep{2008PhRvD..77h4007C}.
These approaches however are not the subject of this work. Instead we operate in
the context of the most commonly employed formalism, the so-called 3+1 formalism
\citep{Alcubierre:2008, 2013rehy.book.....R, 2010nure.book.....B}.

In this formalism, the 4-dimensional spacetime of \ac{GR} is foliated in a
succession of purely spatial hypersurfaces; the \ac{EFE} themselves split in 12
hyperbolic evolution equations, governing the evolution of the fields as time
advances, and 4 elliptic constraint equations. The latter define constraints
that the solution has to satisfy, and at the analytical level are always
satisfied provided the initial data also satisfy them (and as such they must be
solved to generate the initial data itself, see \eg
\citep{2000LRR.....3....5C}). In order to obtain a true solution to the EFE,
these constraints need to be satisfied. Violations may easily lead to unstable
numerical simulations. While the constraints will be always satisfied at the
analytical level, numerical truncation errors will easily cause violations that
can accumulate and destabilize the evolution. It can even be shown that the \adm
\citep{1959PhRv..116.1322A, 1979sgrr.work...83Y} formulation of the \ac{EFE} can
be made strongly hyperbolic by assuming, among other conditions, that the
momentum constraints are identically satisfied \citep{Alcubierre:2008}. These
considerations have motivated the search for alternative, more robust
formulations of Einstein equations. Several approaches have been pursued to
ensure stable numerical evolutions. A widely used and strongly hyperbolic
formulation, namely \bssnok, was introduced in Refs.~\citep{1995PhRvD..52.5428S,
  1998PhRvD..59b4007B, 2009PhRvD..79j4029B, 2004PhRvD..70j4004B}. In this
formulation, the constraint violations cannot be dampened and will accumulate
and grow over time. Despite this shortcoming, it allows for stable, long-term
evolutions, yet in some particularly challenging test cases constraint
violations can grow without bounds, typically crashing the evolution code
\citep{2012PhRvD..85f4040A, 2012PhRvD..85h4004B}.

A simple extension of the \ac{EFE} to include propagating modes for the
constraints, is to generalize a Lagrange multiplier approach, similar to the one
adopted for electrodynamics \citep{Dedner:2002}. The resulting family of
formulations stemming from the Z4 formalism \citep{2003PhRvD..67j4005B}, most
notably \zfourc \citep{2010PhRvD..81h4003B} and \cczfour
\citep{2012PhRvD..85f4040A, 2013PhRvD..88f4049A}, include damping terms designed
with the twofold aim of propagating the constraint violations away from where
they occur and also damping them as they propagate \citep{Gundlach:2005eh}.

It is important to understand that this approach does not guarantee exact
fulfillment of the constraint equations. Techniques to control the growth of
constraint violations however are commonly used in numerical electrodynamics.
Maxwell's equations include conditions such as the absence of magnetic
monopoles, $\nabla\cdot\bfs{B}=0$, which similarly to \ac{GR} are elliptic
equations that the solution of the corresponding evolution equations should
satisfy at all times \citep{Griffiths:2017, Jackson:1998}. An example of a
technique designed to handle these requirements is Dedner's \textit{et al}
method \citep{Dedner:2002}, employed successfully in numerical \ac{MHD} and
\ac{PIC} simulations.

Constraint damping was successfully applied in the first successful merger
simulation \citep{2005PhRvL..95l1101P}, and it has been mainly adopted in
simulations using the \ac{GHG} formulation of the \ac{EFE}
\citep{Pretorius:2006tp,2006CQGra..23S.447L}. One important aspect of the
\ac{GHG} system is that the equations can trivially be recast in first-order
form \citep{2006CQGra..23S.447L}, which is more difficult for \bssnok-like
systems, such as \focczfour \citep{2018PhRvD..97h4053D} or first-order \bssnok
\citep{Brown:2012me}. First-order formulations are particularly important when
solving the \ac{EFE} using finite elements or pseudospectral methods
\citep{Teukolsky:2015ega}, see Refs.~\citep{2018PhRvD..97h4053D},
\citep{Hebert:2018xbk} and \citep{Bhattacharyya:2021dti}.

As recently pointed out, these first-order extensions are subject to additional
curl-constraint, which can render the simulations unstable if not enforced.
Generalizing the idea of divergence cleaning, Ref.~\citep{2020JCoPh.40409088D}
introduced the notion of curl cleaning, which requires to approximately solve
four elliptic equations per constraint (using hyperbolic relaxation), and
applied it to \focczfour. This results in a system with a total of more than a
hundred evolved variables, making the system very expensive to solve and
implement efficiently.

Hence it would be beneficial to have a system of first-order equations that
could be solved using simpler and cheaper approaches. In fact numerical
electrodynamics has benefited also from another class of methods which are able
to maintain a discretized version of the constraints satisfied to machine
accuracy during the evolution, without adding additional equations to the
system. The common feature of these methods is that the electromagnetic
variables are not all defined and stored at the same spatial points in the
computational domain, but on staggered grids. Belonging to this class of methods
are the popular Yee algorithm \citep{Yee66} and \ac{CT} schemes
\citep{Evans1988}, widely used in numerical electrodynamics and \ac{MHD}
simulations.

A constraint preserving scheme for \ac{GR} based on staggered grids was proposed
by Ref.~\citep{Meier:2003bn}. This work identifies as crucial the role played by
the second Bianchi identities in propagating the constraints, and develops a
staggered finite-difference discretization that is able to satisfy them to
machine precision in Riemann normal coordinates. However when such
discretization is applied to general coordinates, the exact fulfillment of the
identities is prevented by the non-cancellation of terms that are cubic in the
Christoffel symbols, which appear as a result of the non-commutativity of
covariant derivatives of the Riemann tensor. As a result the scheme's ability to
exactly propagation of constraints is bounded by the truncation error.

In the present work, we realize the importance of expressing equations as a
system that relates differential forms with the tool of exterior calculus to
obtain discretizations that fulfill the constraints to machine precision, and
apply this idea to obtain a 3+1 formulation of \ac{GR}. Being natural integrands
over submanifolds, differential forms are very well suited to represent
quantities such as total charges inside volumes or fluxes through surfaces. For
this reason, integrating such equations yields a natural discretization that
reflects the geometric properties of the equations themselves, and represents in
a consistent way both the evolution and the constraint equations. Two important
schemes derived from this idea are finite volume and constraint transport
methods. In \ac{MHD}, the former is able to achieve machine precision
conservation of volume-integrated quantities (\eg particle number density) by
locating fluxes at the volume boundaries (cells faces), and the latter is able
to achieve machine precision conservation of surface-integrated magnetic fluxes
(which results in machine-precision fulfillment of $\nabla\cdot\bfs{B}=0$) by
locating electric fields at the surface boundaries (cells edges). I our
endeavor we build upon the fact that a formulation of \ac{GR} in the language
of exterior calculus already exists (in fact it has already been proposed to
exploit it in order to obtain coordinate invariant formulations suitable for
numerical implementation \citep{Frauendiener2006}).

The formulation we develop mirrors at the formal level the equations of
covariant electrodynamics in a moving material medium \citep{Jackson:1998,DPRZ2017}. We
argue that this resemblance would allow to apply the knowledge and the methods
developed in those disciplines to the evolution of dynamical spacetimes; in
particular it would allow to develop \ac{CT} schemes for \ac{NR}, or to apply
divergence- or curl-cleaning methods. In fact, it is conceivable that existing
\ac{MHD} solvers, \eg \citep{Etienne:2015cea,Porth:2016rfi,2020ApJS..249....4S},
could be adapted with minimal effort to solve the equations derived in this work
to evolve dynamical spacetimes instead. This would hold even when adopting
unstructured and moving meshes \citep{Mocz:2014bba}.

This formulation, that we refer to as \dgrem (for
\underline{d}ifferential forms, \underline{g}eneral \underline{r}elativity and
\underline{e}lectro\underline{m}agnetism) also
posses two other desirable features. Firstly, it contains only first order
derivatives in both space and time, which can significantly simplify its
discretization especially with some numerical schemes such as \ac{dG} methods
\citep{Hestaven:2008}. Secondly, it can be written as a system of flux-balance
laws, for the discretization of which a lot of expertise has been amassed over
decades of work \citep{Toro2009}. To the best of the authors' knowledge, no
formulation of the Einstein equations available in the literature combines all
of these advantages.

This work is organized as follows: after defining our notation
(\secref{notation}), in \secref{ext_systems} we introduce our exterior
calculus-based techniques by applying them to the wave equation;
\secref{GR-forms} revisits a formulation of \ac{GR} as a system of equations
written in terms of differential forms. \secref{exploitEM} and
\secref{full-system} are the central part of this work, in which we derive and
present the proposed \dgrem formulation. A summary of the results is given in
\secref{conclusions}, while several appendices provide details of derivations
hinted at in the main text as well as a primer on the theory of exterior
calculus.

\section{Notation and definitions}
\label{sec:notation}

In this section we summarize the notation that is used in the rest of this work,
since due to our reliance on concepts originating from the framework of
exterior calculus, it may not be completely familiar to readers used to the
\ac{NR} literature. We direct the reader to Appendix \ref{sec:diff_forms} and
references therein for more details on differential forms and exterior calculus.
We also collect some definitions used throughout the article, mainly relating to
the 3+1 split of \ac{GR}.

We work within the usual spacetime of general relativity, \ie a 4-dimensional,
Lorentzian, at least twice differentiable manifold $\mathcal{M}$. We
differentiate various type of indices on tensors and differential forms. Letters
from the first half of the Latin alphabet ($a,b,c,\dots$) shall represent, in
any basis, indices ranging from 0 to 3. In a coordinate basis, letters from the
first half of the Greek alphabet ($\alpha,\beta,\gamma,\dots$) shall represent
indices ranging from 0 to 3, and Latin letters from the second half of the
alphabet ($i,j,k,\dots$) shall represent indices ranging from 1 to 3 (\ie
spatial components). The same convention will apply in a non-coordinate
orthonormal basis, but using hatted characters, \ie
$\hat{\alpha},\hat{\beta},\hat{\gamma},\dots$ for indices from 0 to 3, and
$\hat{i},\hat{j},\hat{k},\dots$ for indices from 1 to 3.

In what follows many objects contain non-tensorial indices. These objects are
collections of differential forms, which we also call tensor-valued differential
forms. The indices in these objects simply label the components in the
collection and do not necessarily imply that the collection as a whole transform
a tensor(see \secref{diff_forms} for further details on tensor-valued
differential forms and comments on the terminology). These indices will not be
assigned any particular notation, although their non-tensorial nature will be
indicated in the text.

Without referring to any particular basis, we indicate both tensors and
differential forms with boldface characters; however in the abstract index
notation that we preferentially employ, we drop the boldface font.

We define the following symbols:~\\
\begin{tabular}{ll}
  \toprule\\
  $\eta_{ab}$ & Minkowski metric\\
  $\delta^a_{\ b}$ & Kronecker delta\\
  $\epsilon_{abcd}$ & Levi-Civita symbol\\
  $\varepsilon_{abcd}=\sqrt{-g}\epsilon_{abcd}$ & volume form\\
  $\varepsilon^{abcd}$ & Levi-Civita tensor (dual of volume form)\\
  $\bfs{e}_a$ & vector basis\\
  $\base^a$ & dual basis\\
  $\partial$ & partial derivative\\
  $\nabla$ & covariant derivative\\
  $\ext$ & exterior derivative\\
  $\covext$ & covariant exterior derivative\\
  $\mathcal{L}$ & Lie derivative\\
  $\star$ & Hodge dual\\
  \bottomrule
\end{tabular}~\\~\\
where $g$ denotes the determinant of the metric (see below). Note that all
definitions above, even when written with coordinate basis indices, are valid in
the case of non-coordinate bases too; and that in the definition of basis
vectors and forms, the indices are non-tensorial, simply labeling objects in a
collection.

While the objects we work with are denoted as scalars, vectors, tensors and
differential forms, we actually always mean scalar fields, vector fields, tensor
fields and fields of differential forms respectively, even when this is not
explicitly stated. The same holds for objects that are not tensorial in nature,
such as connection coefficients.

The manifold $\mathcal{M}$ is provided with a metric tensor $g_{\mu\nu}$, for which
we choose the ``mostly plus'' signature $(-, +, +, +)$, and whose determinant is
denoted by $g$. We also summarize here the framework of the 3+1 split of
\ac{GR}, which we employ in order to recast the Einstein equations as an initial
value problem (see standard \ac{NR} textbooks such as \citep{Alcubierre:2008,
  2013rehy.book.....R, 2010nure.book.....B} for more details). We assume that
the spacetime can be foliated in a sequence of tridimensional, purely spatial
hypersurfaces $\Sigma_t$ (\ie the spacetime is assumed to be hyperbolic), each
of which is parametrized by a value of a function $t$. We define the
future-directed unit normal $n_\mu=-\alpha\nabla_\mu t$, where the lapse
function $\alpha$ equals $\alpha=-1/g^{tt}$. From $n_\mu$ we can construct the
metric restricted to each hypersurface $\gamma_{\mu\nu}=g_{\mu\nu}+n_\mu n_\nu$,
which is purely spatial. Considering now the vector $t^\mu=g^{\mu\nu}\nabla_\nu
t$, we identify it with our basis' temporal vector (\ie we choose a basis
adapted to the foliation) and decompose it in a part parallel to $n^\mu$ and one
perpendicular to it: $t^\mu=\bfs{e}_t=\partial_t=\alpha n^\mu+\beta^\mu$. The
purely spatial vector $\beta^\mu$ is called the shift vector. With these
definitions in place we can then state the expressions of $n_\mu$ and
$g_{\mu\nu}$ (or the line element $\dd s$) in a coordinate basis:
\begin{gather*}
  n_\mu=(-\alpha,0,0,0)\quad\textnormal{and}\quad
  n^\mu=\frac{1}{\alpha}(1,-\beta^i)^T\\
  \dd s=-(\alpha^2+\beta_i\beta^i)\dd t^2+2\beta_i\dd t\dd
  x^i+\gamma_{ij}\dd x^i\dd x^j\,,
\end{gather*}
where we have denoted with $x^i$ the spatial coordinates in any hypersurface
$\Sigma_t$ and the $^T$ superscript indicates matrix transposition. We indicate
with $\gamma$ the determinant of $\gamma_{ij}$, $\gamma = \det(\gamma_{ij})$, and note that
$\sqrt{g}=\alpha\sqrt{\gamma}$.

Finally, we define the purely spatial extrinsic curvature
$K_{ij}=-\frac{1}{2}\mathcal{L}_{\bfs{n}}\gamma_{ij}$. As can be surmised
from its definition, the extrinsic curvature is the rate of change of
$\gamma_{ij}$ as measured by an observer moving along $n^\mu$, \ie it is related
to the time derivative of the three-metric $\gamma_{ij}$. We denote its
trace by $K$.

\section{\acp{PDE} in the language of exterior calculus}
\label{sec:ext_systems}

Differential forms are natural integrands on submanifolds, and \acp{PDE} that
can be written as relations between differential forms with the tools of
exterior calculus can be naturally discretized by integration on appropriate
volumes. When such a discretization is applied consistently, the resulting
evolution scheme correctly reflects the geometric structure of the equations. In
turn, this opens up the possibility of developing constraint-preserving evolution
schemes.

In order to introduce the reader to our approach as outlined above, we apply it
in this section to a well-known PDE. Namely, we
explicitly formulate the standard wave equation on a generic spacetime in terms
of differential forms. This helps us setting the stage for reformulating \ac{GR}
and the Einstein equations in the same language in the next section.

\subsection{The wave equation}

Rather than stating the usual wave equation (in terms of scalar or vector fields
and ordinary derivatives) and showing how it can be expressed in terms of
differential forms, we choose here to reverse the exposition order, \ie stating
the equation as a relation between differential forms and then recovering the
usual formulation. This better reflects the derivation the \dgrem formulation of
\ac{GR} in \secref{GR-forms}.

Consider a scalar field (or $0$-form) $\phi$, and its exterior derivative
$\bfs{J}=\ext\phi$ which is of course a $1$-form. $\bfs{J}$ satisfies the
equation
\begin{equation}
  \label{eq:waved}
  -\star^{-1}\ext\star\bfs{J}=0\,.
\end{equation}
Employing the components representation of the exterior derivative and of the
Hodge dual, we can rewrite Eq.~\eqref{eq:waved} as
\begin{equation}
  \varepsilon^{\alpha\beta\gamma\nu}\,
  \partial_{[\nu}\left(\varepsilon_{\alpha\beta\gamma]\mu}\,J^\mu\right)=0 \,.
\end{equation}
Note that in this section we assume for simplicity a coordinate basis, hence the
indices are labeled by Greek letters.

Recalling the definition of $\bfs{\varepsilon}$ it is easy to see that the last
equation becomes
\begin{equation}
  \label{eq:waveJ}
  \frac{1}{\sqrt{-g}}\partial_\mu\left(\sqrt{-g}\,J^\mu\right)=0\,,
\end{equation}
expressing that the divergence of $\bfs{J}$ must vanish. This was to be expected
since operator in \eqref{eq:waved} (sometimes called the codifferential) is a
generalization of the divergence operator (see Eq. \eqref{eq:ddiv}).
Substituting the definition of $\bfs{J}$ as the exterior derivative of $\phi$,
this equation immediately implies
\begin{equation}
  \label{eq:wavephi}
  \frac{1}{\sqrt{-g}}\partial_\mu\left(\sqrt{-g}\,\partial^\mu\phi\right)=0\,,
\end{equation}
\ie the standard homogeneous wave equation for the field $\phi$ in a generic
spacetime.

We now seek too express Eq.~\eqref{eq:wavephi} via a 3+1 formulation, \ie
recasting it as an evolution equation for $\phi$. To this end let us define the
following projections of $\bfs{J}$:
\begin{align}
  \label{eq:waveproj}
  \begin{split}
    \rho &= -n^\mu J_\mu\\
    j_i  &= \gamma^\mu_{\ i}J_\mu\,.
  \end{split}
\end{align}
Substituting these definitions in Eq.~\eqref{eq:waveJ} and recalling the
relationship between the unit normal $n^\mu$, the lapse $\alpha$ and the shift
$\beta^i$, yields the equations
\begin{align}
  \label{eq:wavecons}
  \begin{split}
    \partial_t\left(\sqrt{\gamma}\rho\right)+
    \partial_i\left(\sqrt{\gamma}\rho\mathcal{V}^i\right)&=0\,,\\
    \partial_t j_i+\partial_i\left(\alpha\rho-\beta^kj_k\right)&=0\,,
  \end{split}
\end{align}
where $\mathcal{V}^i = \alpha j^i/\rho - \beta^i$ is the transport velocity of
$\rho$.

These are evolution equations for (quantities related to) the components of
$\bfs{J}$. An evolution equation for $\phi$ itself can easily be recovered from
the definition of $\rho$ and recalling that $J_t=\partial_t\phi$,
resulting in
\begin{equation}
  \partial_t\phi=\alpha\rho+\beta^kj_k\,.
\end{equation}

The wave equation Eq.~\eqref{eq:wavephi}, or the system \eqref{eq:wavecons}, is
subject to a set of differential constraints. Working with differential forms,
this can be seen as follows. The nilpotency of the exterior derivative,
equation \eqref{eq:nilpotent}, immediately gives
\begin{equation}
  \label{eq:waveconstraintd}
  \ext\bfs{J}=\ext\ext\phi=0\,.
\end{equation}

This of course implies that $\star\ext\bfs{J}=0$, and by comparing with
Eq.~\eqref{eq:dcurl}, we can expect this equation to be requiring the curl of
$\bfs{J}$ to vanish. Indeed switching to a components representation and using
the variables $\rho$ and $j_i$, Eq.~\eqref{eq:waveconstraintd} is equivalent to:
\begin{equation}
  \label{eq:waveconstraints}
  \partial_i j_k-\partial_k j_i=0\,.
\end{equation}
These are 3 constraint equations for the spatial components of $\bfs{J}$ (a
fourth equation, stemming from considering the time components and involving the
variable $\rho$, turns out to be identical to the evolution equation for $j_i$).

Eqs.~\eqref{eq:waveconstraints} simply assert the commutativity of second
spatial derivatives of $\phi$, but as the wave equation itself they can be
stated much more compactly and expressively in terms of differential forms.

As mentioned in the introduction, writing the system in terms of differential
forms can be also useful to determine the spatial localization of
variables for a constraint preserving discretization.
However, the direct integration of equations
\eqref{eq:waved} and \eqref{eq:waveconstraintd}, 
would yield a four dimensional discretization 
staggered in time. For methods such as finite-volume,
it is more convenient to derive a semi-discrete
evolution equation with all variables located on the hypersurface $\Sigma_t$.
In order to achieve this, we employ Cartan's ``magic'' formula
(see Eq.~\eqref{eq:cartan_magic} in Appendix \ref{sec:diff_forms}),
and compute the Lie derivative of $\bfs{J}$ and $\star\bfs{J}$,
with respect to the basis vector $\bfs{e}_t$,
which coincides with $\partial_t$.

\begin{align}
\begin{split}
\mathcal{L}_{\bfs{e}_t}\bfs{J}&=\ext(\bfs{e}_t\cdot\bfs{J})\,, \\
\mathcal{L}_{\bfs{e}_t}\star\bfs{J}&=\ext(\bfs{e}_t\cdot\star\bfs{J})\,,
\end{split}
\end{align}
or
\begin{align}
\label{eq:wavecons-forms1}
\partial_t \bfs{J}&=\ext (\alpha \rho + \beta^k j_k)\,, \\
\label{eq:wavecons-forms2}
\partial_t \star \bfs{J}&=\ext\bfs{F}\,,
\end{align}
where the flux form $\bfs{F}$ is defined as
\begin{equation}
    \nonumber
	\bfs{F}=\varepsilon_{ijk} (\alpha j^i - \beta^i )
	 \left(\half \ext x^j \wedge \ext x^k \right) \,.
\end{equation}
The nontrivial components of \eqref{eq:wavecons-forms1}
and \eqref{eq:wavecons-forms2}
give identical equations to those in \eqref{eq:wavecons};
however, the advantage of writing them in this way
is that the submanifolds on which they should be integrated
become explicit. All terms in \eqref{eq:wavecons-forms1} are 1-forms,
and all terms in \eqref{eq:wavecons-forms2} are 3-forms,
which invites to integrate them, respectively, on curves and volumes.
For the purpose of a numerical scheme which decomposes a three-dimentional
simulation domain in zones, this corresponds to integrate the
equations over zone edges and zone volumes.
After applying the Stokes theorem \eqref{eq:stokes},
exterior derivatives are replaced by evaluations of the forms
on zone boundaries (\ie respectively, on zone vertices
and zone faces).

It is straightforward to see that such discretization conserves
globally the volume-integrated `charge' $\rho$: since faces
are shared by two zones, the amount of flux leaving one zone
and entering the other will contribute with opposite signs
to the time update of each zone's content, and the total
charge content in the simulation domain will remain constant
to machine precision as long as there is no flux through the simulation
boundaries.

The discretization also fulfills constraint a discretized version
of equation \eqref{eq:waveconstraints} to machine precision.
This can be seen by integrating equation \eqref{eq:waveconstraintd}
over a zone face (\ie a surface, since it is a 2-form).
The application of Stoke's theorem once more transforms
the exterior derivative into the sum of the forms $\bfs{J}$
integrated on the contour formed by the edges surrounding that face
(\ie the circulation around it).
Also in this case, each of the scalars $\alpha \rho + \beta^k j_k$
defined at zone vertices will be shared by two edges and contribute
to their time update of $\bfs{J}$ with opposite signs,
canceling their contributions to the circulation.
The discretization is therefore able to preserve an integrated
version of constraint \eqref{eq:waveconstraints} to machine
precision when supplied with constraint-fulfilling initial data.

\section{General relativity in the language of exterior calculus}
\label{sec:GR-forms}

In this section, we first lay the groundwork to derive the \dgrem formulation by
outlining a reformulation of the Einstein equations in terms of exterior
calculus and using objects known as the Nester-Witten and Sparling forms. This
results in writing the Sparling equation, which is fully equivalent to the
\ac{EFE}.

We then introduce a change of variables and a particular choice of connection
which ultimately allows us to re-express the Sparling equation, and therefore
the \ac{EFE}, as a system of evolution equations resembling the Maxwell equation
of electrodynamics, \ie the titular \dgrem formulation.

Let's define for convenience the ``hypersurface forms'' as \citep{Szabados1992}:
\begin{equation}
  \label{eq:definition_sigma}
  \bfs{\Sigma}_{a_1\dots a_r}=\frac{1}{(4-r)!}\varepsilon_{a_1\dots a_r a_{r+1}\dots a_4}
  \base^{a_{r+1}}\wedge\dots\wedge\base^{a_4}\,.
\end{equation}
Loosely speaking, they can be thought as (the dual forms to) vectors orthogonal
to submanifolds spanned by given subsets of the basis
$\base^{a_1}\wedge\dots\wedge\base^{a_4}$, \eg the $3$-form
$\Sigma_0=\varepsilon_{0123}\base^1\wedge\base^2\wedge\base^3$ is orthogonal to
the tridimensional hypersurface spanned by $\base^1$, $\base^2$ and $\base^3$.
They satisfy the identity
\begin{equation}
  \label{eq:base_wedge_Sigma}
  \base^b\wedge\bfs{\Sigma}_{a_1\dots a_r}=
  (-1)^{r+1}r\delta^b_{\ [a_1}\bfs{\Sigma}_{a_2\dots a_r]}\,.
\end{equation}

For a manifold with curvature and torsion described, respectively, by the
2-forms $\bfs{\Omega}^a_{\ b}$ and $\bfs{\Xi}^a$, the connection forms
$\cform^a_{\ b}$ (see App. \ref{sec:diff_forms} for a definition) are completely
specified by Cartan's structure equations,
\begin{align}
  \label{eq:CartanI}
  \bfs{\Xi}^a&=\ext\base^a+\cform^a_{\ b}\wedge\base^b\\
  \label{eq:CartanII}
  \bfs{\Omega}^a_{\ b}&=\ext\cform^a_{\ b}+\cform^a_{\ c}\wedge\cform^c_{\ b}\,,
\end{align}
and by the condition of metric compatibility of the connection,
\begin{equation}
  \label{eq:metric-compatibility}
  \ext g_{ab}=\cform_{ab}+\cform_{ba}\,.
\end{equation}
Note that in this last equation the individual components of the metric are seen
as $0$-forms, \ie the metric itself is a tensor-valued $0$-form, hence it is
possible to apply the exterior derivative to it.

The curvature and torsion forms are related to the Riemann and the torsion
tensors $R^a_{\ bcd}$ and $T^a_{\ bc}$ by
\begin{align}
  \bfs{\Omega}^a_{\ b}\wedge\bfs{\Sigma}_{cd}&=R^a_{\ bcd}\;\bfs{\Sigma}\\
  \bfs{\Xi}^a&=T^a_{\ bc}\;\base^b\wedge\base^c\,.
\end{align}

It can be shown \citep{Szabados1992, Frauendiener2006} that the curvature form
is related to the Ricci tensor $R^b_{\ c}$, the curvature scalar
$R=R^b_{\ b}$ and the Einstein tensor
$G^c_{\ d}=R^c_{\ d}-Rg^c_{\ d}$ in the following ways:
\begin{align}
  \label{eq:Curv2RicciTens}
  \bfs{\Omega}^{ab}\wedge\bfs{\Sigma}_{ac}&= R^b_{\ c}\;\bfs{\Sigma}\\
  \label{eq:Curv2RicciScal}
  \bfs{\Omega}^{ab}\wedge\bfs{\Sigma}_{ab}&=R\;\bfs{\Sigma}\\
  \label{eq:Curv2Einstein}
  -\half\bfs{\Omega}^{ab}\wedge\bfs{\Sigma}_{d
    ab}&=G^c_{\ d}\;\bfs{\Sigma}_c\,.
\end{align}

By taking the exterior derivative of Cartan's structure equations
(Eqs.~\ref{eq:CartanI}--\ref{eq:CartanII}), it is possible to obtain the first
and second Bianchi identities,
\begin{align}
  \label{eq:BianchiI}
  \ext\bfs{\Xi}^a&=\bfs{\Omega}^a_{\ e}\wedge\base^e-\cform^a_{\ e}\wedge
  \bfs{\Xi}^e\\
  \label{eq:BianchiII}
  \ext\bfs{\Omega}^a_{\ b}&=\bfs{\Omega}^a_{\ e}\wedge\cform^e_{\ b}-\cform^a_{\ e}
  \wedge\bfs{\Omega}^e_{\ b}\,,
\end{align}
which for a manifold with no torsion and in a coordinate basis take the usual
form
\begin{align}
  \label{eq:BianchiIcoord}
  R_{\mu\alpha\beta\gamma}+R_{\mu\beta\gamma\alpha}+R_{\mu\gamma\alpha\beta}&=0\,,\\
  \label{eq:BianchiIIcoord}
  \nabla_\alpha R_{\mu\nu\beta\gamma}+
  \nabla_\beta R_{\mu\nu\gamma\alpha}+
  \nabla_\gamma R_{\mu\nu\alpha\beta}&=0\,.
\end{align}

To formulate general relativity as a system with exterior derivatives, we first
define a 2-form $\bfs{u}_a$, known as the Nester-Witten form
\citep{Szabados1992, Frauendiener1990, Frauendiener2006}:
\begin{equation}
  \label{eq:nester-witten-definition}
  \bfs{u}_a\coloneqq-\frac{1}{2}\cform^{bc}\wedge\bfs{\Sigma}_{abc}\,.
\end{equation}

Taking its exterior derivative and using the two Cartan structure equations, we
obtain
\begin{equation}
  \label{eq:ext_NW}
  \begin{split}
    \ext \bfs{u}_a =&- \half \bfs{\Omega}^{bc} \wedge \bfs{\Sigma}_{abc} +
    \half \bfs{\Xi}^d \wedge \cform^{bc} \wedge \bfs{\Sigma}_{abcd} \\ &-
    \half \left(\cform^{b}_{\ d} \wedge \cform^{dc} \wedge
    \bfs{\Sigma}_{abc} + \cform^{d}_{\ a} \wedge \cform^{bc} \wedge
    \bfs{\Sigma}_{dbc} \right) \,.
\end{split}
\end{equation}

The terms in parenthesis can be grouped in a 3-form known as the Sparling form:
\begin{equation}
  \label{eq:sparling-form}
  \bfs{t}_a \coloneqq - \half \left(\cform^{b}_{\ d} \wedge
  \cform^{dc} \wedge \bfs{\Sigma}_{abc} + \cform^{d}_{\ a} \wedge
  \cform^{bc} \wedge \bfs{\Sigma}_{dbc} \right) \,,
\end{equation}
whose pull-backs in different basis are related to different expressions for the
gravitational energy-momentum. In particular, in a coordinate basis it is the
Einstein pseudotensor \citep{Frauendiener1990}. For convenience, let us define
$t^b_{\ a}$ such that
\begin{equation}
  \label{eq:tmunu_definition}
  \bfs{t}_a = t^b_{\ a} \bfs{\Sigma}_b\,.
\end{equation}

Assuming no torsion, relation \eqref{eq:Curv2Einstein} and equation
\eqref{eq:ext_NW} can be used to obtain the Sparling equation:
\begin{equation}
  \label{eq:sparling_theorem}
  \ext\bfs{u}_a = \bfs{t}_a + \kappa\bfs{T}_a\,,
\end{equation}
where the non-gravitational energy-momentum 3-form $\bfs{T}_a$ is defined as
\begin{equation}
  \label{eq:energy-momentum_form} 
  \bfs{T}_a = T^\mu_{\ a} \bfs{\Sigma}_\mu \,,
\end{equation}
and where $T^\mu_{\ a}$ are the components of the energy-momentum tensor.

At this point a few comments are necessary. First of all, Eq.
\eqref{eq:sparling_theorem} is equivalent to the Einstein equations
\citep{Szabados1992, Frauendiener1990, Frauendiener2006}, and the sum of the
Nester-Witten and Sparling forms is related to the Einstein tensor by
\begin{equation}
  \ext\bfs{u}_a-\bfs{t}_a=G^b_{\ a}\bfs{\Sigma}_b \,,
\end{equation}
or in components form,
\begin{equation}
  \label{eq:einstein_from_U_t}
  G^c_{\ a}=\frac{1}{\sqrt{-g}}\partial_b\left[\sqrt{-g}
  \,(-\star\bfs{u}_a)^{\ b c}\right]-t^c_{\ a}\,.
\end{equation}

This equivalence holds despite the fact that the index in the objects
$\bfs{u}_a$ and $\bfs{t}_a$ is non-tensorial, \ie the components of the
Nester-Witten form $u_{abc}=(\bfs{u}_a)_{bc}$ \footnote{Here and in the
following, we often employ a simplified notation, writing \eg $u_{abc}$ instead
of the more verbose $(\bfs{u}_a)_{bc}$, when dealing with the components of
various (collections of) differential forms.} are not part of a single 3-indices
tensor, but belong to a collection of four 2-forms labeled by the index $a$,
which transform as ${0\choose2}$-tensors with indices $b$ and $c$ (see also
Appendix \ref{sec:diff_forms}).

This also means that the objects $\bfs{u}_a$ and $\bfs{t}_a$ are not
unique: a different choice of basis 1-forms from which to compute the connection
will lead to different collections of objects, although
Eq.~\eqref{eq:sparling_theorem} will still hold, in the same way as the choice
of different basis and connections does not alter the validity of the Einstein
equations.

Although the non-tensorial behavior of these quantities might be startling, this
behavior is natural, as it is linked to the local flatness of space-time. In the
language of tensors, various quantities (such as the metric first partial
derivatives or energy-momentum pseudotensors) can be made to vanish locally in a
free-falling frame. This is possible owing to the non-tensorial nature of these
objects, as tensors cannot made to vanish by a coordinate (\ie linear)
transformation. By the same token \eg the Sparling form, which is related to
various kinds of energy-momentum pseudotensors \citep{Szabados1992,
  Frauendiener1990}, displays a similar behavior thanks to its own non-tensorial
nature.

\section{Exploiting the analogies with Maxwell's equations}
\label{sec:exploitEM}

\subsection{Evolution equations and constraints}
Equation \eqref{eq:sparling_theorem} presents the Einstein equations as a set of
four equations with a structure very similar to that of the inhomogeneous
Maxwell equations, \ie with the exterior derivative of a 2-form at the left-hand
side and a conserved current at the right-hand side. In fact, taking the
exterior derivative of equation \eqref{eq:sparling_theorem} it can be seen that
the four currents $\bfs{J}_a=\star(\bfs{t}_a+\kappa\bfs{T}_a)$ are globally
conserved. Each antisymmetric tensor $u_{a\mu\nu}$ in the Nester-Witten form
plays the role of the Maxwell 2-form, and in a coordinate basis, equation
\eqref{eq:sparling_theorem} takes a form completely analogous to that
of the inhomogeneous Maxwell equations,

\begin{equation}
    \label{eq:sparling_theorem_coord}
    \partial_b \sqrt{-g}
    \,(-\star u_a^{\ \ b c})
    =\sqrt{-g}(t^c_{\ a} + \kappa T^c_{\ a}) \,.
\end{equation}
Comparing \eqref{eq:sparling_theorem_coord} with \eqref{eq:einstein_from_U_t},
its equivalence to the Einstein equations becomes clear.

Exploiting further the similarity with electrodynamics,
 we can define the following projections of
the Nester-Witten form and its dual
\begin{equation}
  H_a^{\ \mu} \coloneqq u_a^{\ \mu \nu} n_\nu \qquad \text{and} \qquad
  D_a^{\ \mu} \coloneqq - \star u_a^{\ \mu \nu} n_\nu \,.
\end{equation}
This allows to decompose these forms as
\begin{align}
  \label{eq:definition_u}
  u_{a\mu\nu} &= n_\mu H_{a\nu} - n_\nu H_{a\mu} + \varepsilon_{\mu\nu\alpha\beta}
  n^\alpha D_a^{\ \beta}\,, \\
  \label{eq:definition_dualu}
  \star u_{a\mu\nu} &= -n_\mu D_{a\nu} + n_\nu D_{a\mu} +
  \varepsilon_{\mu\nu\alpha\beta} n^\alpha H_a^{\ \beta}\,.
\end{align}
Defining as well the following projections of the components of the Sparling
form and the energy-momentum tensor
\footnote{Note however that these are different from those usually employed in
the literature, where the energy momentum tensor is projected \textit{twice} on
the normal vector and on the hypersurface.},
\begin{align}
  \label{eq:proj_energy-momentum}
  \begin{split}
    \rho_a &\coloneqq n_\mu t^\mu_{\ a} \,,\\[0.5mm]
    s^i_{\ a} &\coloneqq \gamma^i_{\ \mu} t^\mu_{\ a} \,,\\[0.5mm]
    P_a &\coloneqq n_\mu T^\mu_{\ a} \,,\\[0.5mm]
    S^i_{\ a} &\coloneqq \gamma^i_{\ \mu} T^\mu_{\ a} \,.
  \end{split}
\end{align}
Eqs. \eqref{eq:sparling_theorem_coord} can be separated into four constraint
equations
\begin{equation}
  \label{eq:Einstein_constraintsD}
  \mathcal{C}_a\coloneqq
  \partial_i\sqrt{\gamma}D_a^{\ i}-\sqrt{\gamma}(\rho_a+\kappa P_a)=0\,,
\end{equation}
and twelve evolution equations
\begin{equation}
  \label{eq:Einstein_evolD}
  \begin{split}
    &\mathcal{F}^{k}_{\ a}\coloneqq\\
    &\partial_t\sqrt{\gamma}D_a^{\ k}-
    \partial_i\sqrt{\gamma}(\alpha\varepsilon^{kij}H_{aj}+\beta^iD_a^{\ k}-
    \beta^kD_a^{\ i})\\
    &+\sqrt{\gamma}(j^k_{\ a}+\kappa J^k_{\ a})=0
  \end{split}
\end{equation}
where 
\begin{align}
  \label{eq:j_gravitational}
  j^k_{\ a} =& \alpha s^{k}_{\ a} - \beta^k\rho_a,\\
  \label{eq:J_matter}
  J^k_{\ a} =& \alpha S^{k}_{\ a} - \beta^k P_a \,.
\end{align}

The fulfillment of equations \eqref{eq:Einstein_constraintsD} is equivalent to
that of the Einstein constraints. This can be seen by the definition of the
usual Hamiltonian and momentum constraints and the 3+1 evolution equations
\citep{Frittelli1997} as
\begin{align}
  \begin{split}
	\mathcal{H}&\coloneqq n^\mu n^\nu(G_{\mu \nu}-\kappa T_{\mu \nu})=0\,,\\
	\mathcal{M}_i&\coloneqq \gamma^\mu_{\ i}n^\nu (G_{\mu \nu}-\kappa T_{\mu \nu})=0\,,\\
	\mathcal{E}_{ij}&\coloneqq \gamma^\mu_{\ i}\gamma^\nu_{\ j}(G_{\mu \nu}-\kappa T_{\mu
      \nu})=0\,,
  \end{split}
\end{align}
from which
\begin{align}
  \begin{split}
    \mathcal{C}_0&=-\mathcal{H}\,,\\
    \mathcal{C}_i&=-\mathcal{M}_i/\alpha\,,\\
    \mathcal{F}^{i}_{\ 0}&=\alpha\mathcal{M}^i+\beta^i\mathcal{H}\,,\\
    \mathcal{F}^{i}_{\ j}&=\mathcal{E}^i_{\ j}+\beta^i\mathcal{M}_j/\alpha\,,
  \end{split}
\end{align}
and therefore $\mathcal{C}_a=0$ is equivalent to $\mathcal{M}_i=0$ and
$\mathcal{H}=0$. The twice-contracted second Bianchi identities imply that if
the Hamiltonian constraint is fulfilled on a space-like hypersurface, its
fulfillment on the ``next'' hypersurface is guaranteed as long as the momentum
constraints are satisfied exactly and the system is evolved using evolution 3+1
Einstein equations \citep{Frittelli1997}. Similar equations for the propagation
of constraints $\mathcal{C}_a$ can be obtained after taking the exterior
derivative of the Sparling equation \eqref{eq:sparling_theorem}. This results in
a set of equations equivalent to the twice-contracted second Bianchi identities,
of the form
\begin{equation}
  \partial_t\sqrt{-g}\mathcal{C}_a+\partial_i\sqrt{-g}\mathcal{F}^i_{\ a}=0\,.
\end{equation}

Therefore, also in this case the evolution equations for $D_k^{\ i}$ and the
exact fulfillment of the momentum constraints $\mathcal{C}_i$ are sufficient to
propagate the fulfillment of $\mathcal{C}_0$ between subsequent hypersurfaces.

\subsection{Energy-momentum conservation}

The exterior derivative of equation \eqref{eq:sparling_theorem} can also be used
to obtain evolution equations for the ``charge densities'' $\rho_a$ and $P_a$,
as it expresses the global conservation of the sum of their currents,
\begin{equation}
  \ext (\bfs{t}_a + \kappa \bfs{T}_a) =0\,.
\end{equation}
Together with the local conservation of matter energy-momentum
$\covext\bfs{T}_a=0$\footnote{In this equation $\covext$ represents the exterior
covariant derivative (see Appendix \ref{sec:diff_forms}), and the equation is
equivalent to the usual $\nabla_\mu T^{\mu\nu}=0$.}, this gives
\begin{align}
  \ext \bfs{T}_a &= \cform^b_{\ a} \wedge \bfs{T}_b \qquad
  \text{and} \label{eq:T_cons}\\
  \ext \bfs{t}_a &= -\kappa \ \cform^b_{\ a} \wedge \bfs{T}_b\,,
  \label{eq:t_cons}
\end{align}
or in component form and in a coordinate basis,
\begin{align}
  \partial_\mu \sqrt{-g}\ T^\mu_{\ a} &= \sqrt{-g}\ \omega^b_{\ a\mu}
  T^\mu_{\ b}\label{eq:em_cons} \qquad \text{and} \\
  \partial_\mu \sqrt{-g}\ t^\mu_{\ a} &= -\kappa\ \sqrt{-g}\ \omega^b_{\ a\mu}
  T^\mu_{\ b}\,.
\end{align}
Substituting the projections defined above (Eq.~\ref{eq:proj_energy-momentum}),
\begin{align}
  \label{eq:grav-energy-momentum-cons}
  \partial_t \sqrt{\gamma} \rho_a + \partial_i \sqrt{\gamma} (\alpha s^i_{\ a} -
  \beta^i \rho_a) &= -\kappa \sqrt{\gamma} Q_a \,,\\
  \label{eq:matt-energy-momentum-cons}
  \partial_t \sqrt{\gamma} P_a + \partial_i \sqrt{\gamma} (\alpha S^i_{\ a} -
  \beta^i P_a) &= \ \sqrt{\gamma}Q_a \,,
\end{align}
where
\begin{equation}
  \label{eq:Q}
  Q_a= -(\omega^b_{\ at} + \omega^b_{\ ai}\beta^i)P_b + \alpha \omega^b_{\ ai}
  S^i_{\ b} \,.
\end{equation}

The physical interpretation of Eqs. \eqref{eq:sparling_theorem},
\eqref{eq:T_cons} and \eqref{eq:t_cons} can be that of four vector fields
described by the four 2-forms $\bfs{u}_a$ which have as sources two currents
$\star \kappa\bfs{T}_a$ and $\star \bfs{t}_a$. The sum of the latter two is
globally conserved, but they exchange charge (in this case, energy and momentum)
via the ``force'' term $\kappa \cform^b_{\ a} \wedge \bfs{T}_b$. These currents
are those of gravitational ($\star \bfs{t}_a$) and non-gravitational ($\star
\kappa\bfs{T}_a$) energy and momentum. Eqs. \eqref{eq:Einstein_constraintsD} and
\eqref{eq:Einstein_evolD} are the analogue of the inhomogeneous Maxwell
equations in 3+1 form, and equations \eqref{eq:grav-energy-momentum-cons} and
\eqref{eq:matt-energy-momentum-cons} that of the conservation of the two
charges.

While Eqs.~\eqref{eq:grav-energy-momentum-cons} and
\eqref{eq:matt-energy-momentum-cons} convey an interesting physical picture of
energy exchange between the purely gravitational and the matter sector, there is
another possibility of how to read these equations in practice. Adding up
\eqref{eq:grav-energy-momentum-cons} and \eqref{eq:matt-energy-momentum-cons}, we
obtain
\begin{align}
  \label{eq:rhoP_tot}
  &\partial_t \left[ \sqrt{\gamma} \left( \rho_a + \kappa P_a \right)
    \right]\nonumber\\
  + &\partial_i\left[ \sqrt{\gamma} \left( \alpha \left( s^i_{\ a} +
    \kappa S^i_{\ a} \right) - \beta^i \left( \rho_a + \kappa P_a \right)
  \right) \right]=0\,.
\end{align}
When comparing this equation with the equation of energy-momentum conservation
\eqref{eq:em_cons}, it is striking to see that using the Sparling form all
source terms in \eqref{eq:rhoP_tot} have disappeared. In this formulation, the
geometric source terms of Eq. \eqref{eq:em_cons} have been recast into a fully
flux conservative form. A similar observation has recently also been made by
\citet{Clough:2021qlv}. While previously such a formulation was known to exist
for the time-component of Eq. \eqref{eq:em_cons} in static spacetimes
\citep{Gammie:2003rj}, this is the case here in any dynamical and non-dynamical
spacetime. While sounding trivial at first, such a formulation opens up the
exciting prospects of applying advanced techniques from flux-balance equations
to the Einstein-Matter system, such first-order flux limiting
\citep{2013JCoPh.242..169H} to ensure positivity of energy- and momentum
densities.

This is particularly interesting when combined with the relativistic (magneto-)
hydrodynamics description of the matter part, for which non-trivial constraints
on the physicality of the energy-momentum density $P_a$ exist. A formulation
such as this one, clearly separating gravitational and matter contribution, as
well as having no explicit sources, might make it possible to transfer advances
made on physicality preserving schemes in special relativity over to general
spacetimes \citep{2017PhRvD..95j3001W,2018ZaMP...69...84W}.

\subsection{Choosing a connection}
\label{sec:connection}

In \secref{GR-forms}, we showed that the Einstein equations and
the conservation of energy and momentum can be expressed as a system of
equations with close similarities to the inhomogeneous Maxwell equations and the
equation of charge conservation. However, even assuming that we have equations
to evolve the matter energy-momentum, in order to close the system we need to
specify a way of updating the quantities that appear in the equations for which
no evolution equation is provided, that is, $\sqrt{\gamma}$, $\omega^b_{\ ac}$,
$H_a^{\ \mu}$, and $s^\mu_{\ a}$. To find relations between these quantities and
the evolved variables, we start by noticing that the Hodge dual of the
Nester-Witten form can be written in terms of the connection as
\begin{equation}
  \label{eq:dual_u_from_omega}
  (\star\bfs{u}_a)^{bc}=\omega^{[bc]}_{\ \ \ a}
  +\delta^b_{\ a}\omega^{[cd]}_{\ \ \ d}
  -\delta^c_{\ a}\omega^{[bd]}_{\ \ \ d}\,.
\end{equation}
The detailed calculation is provided in Appendix \secref{omega_to_dualU}.
Relation \eqref{eq:dual_u_from_omega} can be contracted to obtain
\begin{equation}
  \label{eq:u_omega_contraction}
  (\star\bfs{u}_c)^{bc}=-2\omega^{[bc]}_{\ \ \ c}\,,
\end{equation}
from which
\begin{equation}
  \label{eq:omega_from_dual_u}
  \omega^{[bc]}_{\ \ \ a}=(\star\bfs{u}_a)^{bc}
  -\half\delta^b_{\ a}(\star\bfs{u}_d)^{cd}
  +\half\delta^c_{\ a}(\star\bfs{u}_d)^{bd}\,.
\end{equation}
This shows that the part of the connection that is antisymmetric with
respect to its first two indices is completely determined by the Nester-Witten
form. Since the full connection appears in other parts of the system, namely
inside $\bfs{t}_a$ (Eq.~\eqref{eq:sparling-form}) and $Q_a$ (Eq.~\eqref{eq:Q}),
in principle it could be necessary to evolve also the part that is symmetric
with respect to these indices. To simplify calculations, it would be useful to
exploit the non-uniqueness of the Nester-Witten and the Sparling forms to build
them from a connection that is purely antisymmetric with respect to its first
two indices. This is the case for the 
spin connection \citep[c.f. Appendix J of][]{2004sgig.book.....C},
also known as the Ricci rotation coefficients
\citep[c.f. Section 3.4b of][]{1984ucp..book.....W}.
For an orthonormal vector basis
$\{\bfs{e}_{\hat{\alpha}}\}$ with dual 1-form basis
$\{\base^{\hat{\alpha}}\}$, the spin connection
$\cform^{\hat{\alpha}}_{\ \hat{\beta}}
=\omega^{\hat{\alpha}}_{\ \hat{\beta}\hat{\mu}}\base^{\hat{\mu}}$
is defined by

\begin{equation}
	\partial_{\hat{\nu}}\bfs{e}_{\hat{\mu}}
	\coloneqq A^{\nu}_{\ \hat{\nu}} \partial_{\nu}\bfs{e}_{\hat{\mu}}=
	\omega^{\hat{\alpha}}_{\ \hat{\mu}\hat{\nu}}\base^{\hat{\mu}} \bfs{e}_{\hat{\alpha}} \,,
\end{equation}
where $A^{\nu}_{\ \hat{\nu}}$ are the coefficients that
relate the orthonormal basis to the coordinate basis $\{\bfs{e}_{\alpha}\}$,
$\bfs{e}_{\hat{\nu}}=A^{\nu}_{\ \hat{\nu}} \bfs{e}_{\nu}$.
The orthonormal 1-form basis $\{\base^{\hat{\mu}}\}$ and the coordinate basis
$\{\base^{\mu}\}$
are related by the transformations
$\base^{\hat{\mu}}=A^{\hat{\mu}}_{\ \mu}\base^{\mu}$ and
$\base^{\mu}=A^{\mu}_{\ \hat{\mu}}\base^{\hat{\mu}}$. 
The form of the metric when expressed in an orthonormal basis is that
of Minkowski metric, and is therefore constant.
From metric compatibility \eqref{eq:metric-compatibility}, it follows
that this connection is completely antisymmetric with respect to its first
two indices. This can also be seen from the metricity condition,
which states that the covariant derivative of the metric must vanish,

\begin{equation}
	\nabla_{\hat{\alpha}} \eta_{\hat{\mu}\hat{\nu}}
	= \partial_{\hat{\alpha}} \eta_{\hat{\mu}\hat{\nu}}
	  -\omega^{\hat{\beta}}_{\ \hat{\mu}\hat{\alpha}} \eta_{\hat{\beta}\hat{\nu}}
  	  -\omega^{\hat{\beta}}_{\ \hat{\nu}\hat{\alpha}} \eta_{\hat{\mu}\hat{\beta}}=0\,.
\end{equation}

In what follows, we still express the equations in a coordinate basis to keep
the convenience of directly integrating $p$-forms over coordinate submanifolds,
but construct an orthonormal tetrad field to obtain the connection from which
$\bfs{u}_a$ and $\bfs{t}_a$ are defined.

Given the 3+1 foliation of
the spacetime, a natural choice for the tetrad is that of an Eulerian observer
moving at velocity $n^\mu$, \ie we take the vector $n^\mu$ to be part of the
basis we are seeking. In order to accomplish this, the components of the tetrad
basis one-forms in the coordinate basis can be written as:
\begin{align}
  \label{eq:tetrad_basis}
  A^{\hat{0}}_{\ \mu}&=(\alpha,0_i)=-n_\mu,\\
  A^{\hat{i}}_{\ \mu}&=(\beta^{\hat{i}},A^{\hat{i}}_{\ i}) ,
\end{align}
where
\begin{align}
  \eta_{\hat{\mu}\hat{\nu}}A^{\hat{\mu}}_{\ \mu} A^{\hat{\nu}}_{\ \nu}&=g_{\mu\nu},\\
  \beta^{\hat{i}}&=A^{\hat{i}}_{\ i}\beta^i,\\
  \delta_{\hat{i}\hat{j}}A^{\hat{i}}_{\ i}A^{\hat{j}}_{\ j}&=\gamma_{ij}\,.
\end{align}
Conversely, the inverse transformation is given by
\begin{align}
  \label{eq:tetrad_basis_inv}
  A^{\mu}_{\ \hat{0}} & = (1/\alpha,-\beta^i/\alpha) = n^\mu\\
  A^{\mu}_{\ \hat{i}} & = \left( \begin{array}{c}
  0_{\hat{i}}	\\
  A^i_{\ \hat{i}}	
  \end{array}
	\right)  \,,
\end{align}
where also
\begin{align}
  \eta^{\hat{\mu}\hat{\nu}}A_{\ \hat{\mu}}^\mu A_{\ \hat{\nu}}^\nu &= g^{\mu\nu}\\
  \delta^{\hat{i}\hat{j}}A_{\ \hat{i}}^i A_{\ \hat{j}}^j &= \gamma^{ij}\,.
\end{align}

The spin connection is calculated from the commutation coefficients of the
basis, $c^{\hat{\alpha}}_{\ \hat{\mu}\hat{\nu}}$, which in turn can be obtained
either as the commutators of the basis vectors, or as the exterior derivatives
of the basis 1-forms. 
While the two quantities coincide when expressed in the
orthonormal basis, they obey different transformation laws, transforming,
respectively, as a vector and as a 2-form. To keep exploiting the analogies with
electromagnetism, we decide to calculate the commutation coefficients in the
second way, and define the set of 2-forms
\begin{equation}
  \label{eq:F_definition}
  \bfs{F}^{\hat{\alpha}} = \ext \base^{\hat{\alpha}}\,,
\end{equation}
which in a coordinate basis takes the form
\begin{equation}
  \label{eq:F_definition_coordinate}
        {F}^{\hat{\alpha}}_{\ \mu\nu} =
        \partial_\mu A^{\hat{\alpha}}_{\ \nu}
        - \partial_\nu A^{\hat{\alpha}}_{\ \mu} \,.
\end{equation}

The commutation coefficients are equal to the components of these forms when
expressed in the tetrad basis,
\begin{equation}
  \label{eq:comm_coeff}
  c^{\hat{\alpha}}_{\ \hat{\mu}\hat{\nu}}={F}^{\hat{\alpha}}_{\ \hat{\mu}\hat{\nu}}=
  A_{\ \hat{\mu}}^\mu A_{\ \hat{\nu}}^\nu{F}^{\hat{\alpha}}_{\ \mu\nu}\,,
\end{equation}
and the connection can be calculated as
\begin{align}
  \label{eq:Ricci_connection}
    \omega_{\hat{\alpha}\hat{\mu}\hat{\nu}} & = 
    \half \left( c_{\hat{\mu}\hat{\alpha}\hat{\nu}}
    + c_{\hat{\nu}\hat{\alpha}\hat{\mu}}
    - c_{\hat{\alpha}\hat{\mu}\hat{\nu}} \right) \,.
\end{align}
The striking similarity of the spin connection to the Levi-Civita
connection is by no means a coincidence. 
The spin connection can be used to generalize the covariant
derivate for general tensors $V_{\hat{\alpha}}^\nu$,
\begin{align}
  D_\mu V^\nu_{\hat{\alpha}} = \partial_\mu V^\nu_{\hat{\alpha}} +
  \Gamma^{\nu}_{\mu\beta} V^\beta_{\hat{\alpha}} -
  \omega^{\hat{\gamma}}_{\hat{\alpha}\mu} V^\nu_{\hat{\gamma}}\,.
  \label{eq:Dcov_general}
\end{align}
It can be shown that this derivative is covariant in the tetrad and the
coordinate frame.
The specific form of the spin connection  \eqref{eq:comm_coeff} now
arises because the choice of 2-forms in \eqref{eq:F_definition} is
equivalent to demanding metric compatibility of the local flat metric in
the tetrad under transformations of the generalized covariant derivative
\eqref{eq:Dcov_general},
\begin{align}
  D_{{\mu}} \eta_{\hat{\alpha}\hat{\beta}} \,=\, 0\,.
  \label{eq:Deta}
\end{align}
In the same way, that metric compatibility of the space-time metric
uniquely results in the Levi-Civita connection, the choice of
\eqref{eq:Deta} imposes the form of the connection coefficients
\eqref{eq:comm_coeff}. Put differently, we have defined both the global
manifold and the local tetrad to be torsion free.

Conversely, by inverting relation \eqref{eq:Ricci_connection} it can be found
that the forms $\bfs{F}_{\hat{\alpha}}$ collect the antisymmetric part of the spin
connection with respect to the last two indices,
\begin{equation}
  \label{eq:F_from_connection}
  F_{\hat{\alpha}\hat{\mu}\hat{\nu}}=
  \omega_{\hat{\alpha}\hat{\nu}\hat{\mu}}-
  \omega_{\hat{\alpha}\hat{\mu}\hat{\nu}}\,.
\end{equation}

We now define the following projections of $\bfs{F}_{\hat{\alpha}}$ and its dual
$\star\bfs{F}_{\hat{\alpha}}$ as
\begin{align}
  \label{eq:EBprojections}
  \begin{split}
    E^{\hat{\alpha} \mu}&\coloneqq F^{\hat{\alpha} \mu \nu}n_\nu\\
    B^{\hat{\alpha} \mu}&\coloneqq \star F^{\hat{\alpha} \mu \nu}n_\nu\,,
  \end{split}
\end{align}
so that we can write their components as
\begin{align}
  \label{eq:F_components}
  \begin{split}
    F^{\hat{\alpha}}_{\ \mu \nu} &= n_\mu E^{\hat{\alpha}}_{\ \nu} - n_\nu
    E^{\hat{\alpha}}_{\ \mu} - \varepsilon_{\mu\nu\lambda\sigma} n^\lambda
    B^{\hat{\alpha} \sigma}\\
    \star F^{\hat{\alpha}}_{\ \mu \nu} &= n_\mu B^{\hat{\alpha}}_{\ \nu} - n_\nu
    B^{\hat{\alpha}}_{\ \mu} + \varepsilon_{\mu\nu\lambda\sigma} n^\lambda
    E^{\hat{\alpha} \sigma}\,.
  \end{split}
\end{align}
Substituting equations \eqref{eq:F_components} and \eqref{eq:tetrad_basis} in
\eqref{eq:F_definition_coordinate}, we find the following evolution equations
for the transformation coefficients on the slice
\begin{equation}
  \label{eq:Aupdate}
  \begin{split}
    \partial_t A^{\hat{i}}_{\ i}-\partial_i\beta^{\hat{i}}
    &=-\alpha E^{\hat{i}}_{\ i}+\varepsilon_{ilk}\beta^lB^{\hat{i} k}\\
    &=-\alpha E^{\hat{i}}_{\ i}+\sqrt{\gamma}\epsilon_{ilk}\beta^l B^{\hat{i} k}\,,
  \end{split}
\end{equation}
along with the constraints
\begin{align}
  \label{eq:EBconstraints}
  \begin{split}
    E^{\hat{0}}_{\ i}&=-\partial_i\ln\alpha,\\
    \sqrt{\gamma}B^{\hat{0}i}&=0,\\
    \sqrt{\gamma}B^{\hat{i}k}&=\epsilon^{ijk}\partial_iA^{\hat{i}}_{\ j}\,.
  \end{split}
\end{align}
These equations are in close analogy to electromagnetism, with the role of the
3-vector potential played by $A^{\hat{i}}_{\ i}$ and that of the scalar potential
played by $\beta^{\hat{i}}$. It is interesting to see that equation
\eqref{eq:F_definition} does not provide evolution equations for $\alpha$ and
$\beta^i$, which is in agreement with the gauge freedom of the spacetime
foliation.

By taking the exterior derivative of equation \eqref{eq:F_definition}, we obtain
\begin{equation}
  \label{eq:dF}
  \ext \bfs{F}^{\hat{\alpha}} = 0\,,
\end{equation}
which is nothing more than the first Bianchi identity, as can be seen by
comparing equation \eqref{eq:F_definition} with \eqref{eq:CartanI} and
\eqref{eq:BianchiI} with \eqref{eq:dF}. Using the projections in definition
\eqref{eq:EBprojections}, equation \eqref{eq:dF} splits in four equations with a
form analogous to the Gauss law for magnetism, namely
\begin{equation}
  \label{eq:BianchiDivB}
  \partial_i \sqrt{\gamma} B^{\hat{\alpha}i} = 0 \,,
\end{equation}
and twelve evolution equations analogous to the Faraday equation,
\begin{align}
  \label{eq:BianchiFaraday}
  \begin{split}
    &\partial_t\sqrt{\gamma}B^{\hat{\alpha}k}\\
    &+\partial_i\sqrt{\gamma}(\alpha\epsilon^{ijk}E^{\hat{\alpha}}_{\ j}
    -\beta^iB^{\hat{\alpha}k}+\beta^kB^{\hat{\alpha}i})=0\,.
  \end{split}
\end{align}
For $\hat{\alpha}=\hat{0}$, equations \eqref{eq:BianchiDivB} and
\eqref{eq:BianchiFaraday} are trivially fulfilled, since $B^{\hat{0}i}=0$, and
equation \eqref{eq:BianchiFaraday} becomes simply an expression of the
commutativity of the partial derivatives of $\alpha$.

\subsection{Closing the system}
\label{sec:closing_system}

We have now obtained all the evolution equations of the system, and can list the
elements of the state vector as
$\vec{U}=\{A^{\hat{k}}_{\ i},D_{\hat{\alpha}}^{\ i},\rho_{\hat{\alpha}},P_{\hat{\alpha}}\}$,
where the first 25 quantities determine the state of the gravitational field,
while the four momentum densities $P_{\hat{\alpha}}$ depend on the properties of
matter. Additionally, we need a set of relations to obtain the remaining quantities that
appear in their evolution equations, namely
$\vec{Q}=\{\sqrt{\gamma},E^{\hat{\alpha}}_{\ i},B^{\hat{k}i},
H_{\hat{\alpha}}^{\ i},s^i_{\ \hat{\alpha}},S^i_{\ \hat{\alpha}},Q_{\hat{\alpha}}\}$, where again
the momentum fluxes $S^i_{\ \hat{\alpha}}$ depend on the properties of matter.
Although $\sqrt{\gamma}$ and $B^{\hat{k}i}$ can in principle be obtained as the
determinant and the curl of $A^{\hat{k}}_{\ i}$, respectively, it may be useful
to evolve them with an independent evolution equation. In the case of
$\sqrt{\gamma}$, the reason being to evolve it at the side of conformally
rescaled quantities or to avoid errors associated to the numerical computation
of the determinant.
An evolution equation for $\sqrt{\gamma}$,
can be obtained by using \eqref{eq:definition_sigma} to define
the hypersurface form orthogonal to $-n_\mu$, that is, to $\base^{\hat{0}}$,
and taking its exterior derivative. The resulting expression has the form
of a conservation equation for volume,
\begin{equation}
    \label{eq:sqrtgammaevol}
	\partial_t \sqrt{\gamma} -\partial_i \sqrt{\gamma} \beta^i
	= \frac{5}{4}\sqrt{\gamma} D_{\hat{k}}^{\ \hat{k}} \,,
\end{equation}
in which the rate of change in volume of a small region
is related to the amount of volume that enters through its
boundaries due to the motion of coordinates (represented by $\beta^i$)
plus the amount of volume generated within the region due to the
presence of a field $D_{\hat{k}}^{\ i}$. A derivation of this equation
can be found in Appendix \ref{sec:dt_sqrtgamma_derivation}.

In the case of $B^{\hat{k}i}$, an independent evolution equation
\eqref{eq:BianchiFaraday} may be needed in constraint-damping schemes (as
opposed to constrained transport schemes), where the identity of $B^{\hat{k}i}$
as the curl of $A^{\hat{k}}_{\ i}$, and therefore the fulfillment of the first
Bianchi identity, is not guaranteed and needs to be enforced. The gauge
functions $\vec{G}=\{\alpha,\beta^{i}\}$ may belong to either of the sets
$\vec{U}$ or $\vec{Q}$,
depending on whether we enforce new differential equations for their evolution,
or set them as algebraic functions of $\vec{U}$. Finally, the rest of quantities
can be obtained from algebraic relations analogous to the constitutive equations
in electrodynamics.

These constitutive relations can be obtained from equations
\eqref{eq:omega_from_dual_u} and \eqref{eq:Ricci_connection}, which determine
the relations between the connection coefficients in terms of the Nester-Witten
form and the form $\bfs{F}^{\hat{\alpha}}$ in the orthonormal frame.
\begin{align}
  \label{eq:D0_EB}
  D_{\hat{0}}^{\ \hat{i}} &= -\epsilon^{\hat{i}\hat{j}\hat{k}} B_{\hat{j} \hat{k}}, \\
  \label{eq:Dk_EB}
  D_{\hat{k}}^{\ \hat{i}} &= -\half (E_{\hat{k}}^{\ \hat{i}} + E^{\hat{i}}_{\ \hat{k}})
  + \delta_{\hat{k}}^{\ \hat{i}} E^{\hat{l}}_{\ \hat{l}} ,\\
  \label{eq:H0_EB}
  H_{\hat{0} \hat{i}} &= \half \epsilon_{\hat{i}\hat{j}\hat{k}} E^{\hat{j}\hat{k}}, \\
  \label{eq:Hk_EB}
  H_{\hat{k} \hat{i}} &= - B_{\hat{i} \hat{k}}
  + \half \delta_{\hat{k} \hat{i}} B_{\hat{l}}^{\ \hat{l}}
  - \epsilon_{\hat{k}\hat{i}\hat{l}} E_{\hat{0}}^{\ \hat{l}} , \\
  \label{eq:E0_DH}
  E^{\hat{0}}_{\ \hat{i}} &= \frac{3}{2} D_{\hat{0}\hat{i}} - \frac{1}{2}
  \epsilon_{\hat{i}\hat{j}\hat{k}} H^{\hat{j} \hat{k}} ,\\
  \label{eq:Ek_DH}
  E^{\hat{j}}_{\ \hat{k}} &= - D_{\hat{k}}^{\ \hat{j}} -\half
  \delta^{\hat{j}}_{\ \hat{k}} D_{\hat{l}}^{\ \hat{l}}
  +\epsilon^{\hat{j}}_{\ \hat{k}\hat{l}} H_{\hat{0}}^{\ \hat{l}} ,\\
  \label{eq:B0_DH}
  B^{\hat{0}\hat{i}} &= -\epsilon^{\hat{i}\hat{j}\hat{k}} D_{\hat{j}\hat{k}} , \\
  \label{eq:Bk_DH}
  B^{\hat{i}\hat{j}} &= \delta^{\hat{i}\hat{j}} H_{\hat{l}}^{\ \hat{l}}
  - H^{\hat{i}\hat{j}} 
  + \half \left( H^{\hat{j}\hat{i}}-H^{\hat{i}\hat{j}}
  + \epsilon^{\hat{i}\hat{j}\hat{k}} D_{\hat{0} \hat{k}} \right) \,.
\end{align}

We are interested in obtaining the unknown quantities ($E^{\hat{k}}_{\ \hat{i}}$ and 
$H_{\hat{\alpha}}^{\ \hat{i}}$) needed for evolution
 from the known
evolved variables ($D_{\hat{\alpha}\hat{i}}$ and $B^{\hat{k}\hat{i}}$). We have
already expressions for $B^{\hat{0}\hat{i}}$, and $E^{\hat{0}}_{\ \hat{i}}$,
since they are determined by the gauge from Eqs.~\eqref{eq:EBconstraints}.
Therefore, the required relations are given by Eqs.~\eqref{eq:Ek_DH} and
\eqref{eq:Hk_EB}.

The the system \eqref{eq:D0_EB}-\eqref{eq:Hk_EB} and
\eqref{eq:E0_DH}-\eqref{eq:Bk_DH}, also gives constraints on some of the
variables determined by evolution. In particular, Eqs.~\eqref{eq:D0_EB} and
\eqref{eq:Dk_EB} imply that the $D_{\hat{i}\hat{j}}$ is symmetric, and that
$D_{\hat{0}}^{\ \hat{k}}$ is related to the anti-symmetric part of
$B_{\hat{i}\hat{j}}$. This is a consequence of the symmetry of the Einstein
equations, which allow to express some of the quantities as linear combinations
of the others. In principle this could help us reducing the number of necessary
evolution equations, as one could evolve just $D_{\hat{i}\geq\hat{j}}$ and
$B_{\hat{i}\hat{j}}$, and obtain their derived quantities when they are needed.
However, the variables involved in these constraints have different geometric
meanings. For example, $D_{\hat{i}}^{\ \hat{j}}$ is $j$-th component of the
3-vector field $\bfs{D}_{\hat{i}}$, while $D_{\hat{j}}^{\ \hat{i}}$ is the
$i$-th component of $\bfs{D}_{\hat{j}}$, and they are orthogonal to different
surfaces. This will become relevant when designing a staggered scheme that
allows to keep the constraints fulfilled to machine precision, and where
$D_{\hat{i}}^{\ \hat{j}}$ and $D_{\hat{j}}^{\ \hat{i}}$ will have different
spatial representations, so it may be convenient to evolve them separately. The
case of $D_{\hat{0}}^{\ \hat{k}}$ is slightly different, since the propagation
of constraint $\mathcal{C}_0$ is ensured by the exact fulfillment of
$\mathcal{C}_i$, so it might be possible to drop completely its evolution as
well as that of the gravitational energy $\rho_{\hat{0}}$. An approximate value
of $D_{\hat{0}}^{\ \hat{k}}$ can then always be obtained from
$B_{\hat{i}\hat{j}}$ and an approximate value of $\rho_{\hat{0}}$ from
calculating the divergence of $D_{\hat{0}}^{\ \hat{k}}$ and taking the
difference with the matter energy $P_{\hat{0}}$ according to equations
\eqref{eq:Einstein_constraintsD}. However, their evolution can still be useful
to keep track of the transport of gravitational energy and to provide
information on the differences between the components of $B_{\hat{i}\hat{j}}$,
which might increase the accuracy of interpolations.

Finally, another interesting feature of the constitutive relations
\eqref{eq:D0_EB}-\eqref{eq:Bk_DH} is that they provide no means of calculating
$H_{\hat{0}\hat{i}}$ from the evolved variables. Similarly as for the gauge
variables $\alpha$ and $\beta^i$, this indicates that $H_{\hat{0}\hat{i}}$
represents an additional freedom of the formulation, and in fact, it can be
related to the custom choice of rotating the tetrad bases between different
hypersurfaces. To see this, let us consider a special case of a spacetime devoid
of matter and gravitational energy-momentum, for which
$D_{\hat{\alpha}}^{\ i}=0$ is a solution to constraints
\eqref{eq:Einstein_constraintsD}. Choosing a gauge in which the shift is zero
and the lapse is one (geodesic gauge), the evolution equations for the tetrad
coefficients (Eq.~\ref{eq:Aupdate}) read
\begin{equation}
  \partial_t A^{\hat {i}}_{\ i} = -E^{\hat {i}}_{\ i} = -A^{\hat {j}}_{\ i}
  \epsilon^{\hat{i}}_{\ \hat{j}\hat{k}} H_{\hat{0}}^{\ \hat{k}} \,,
\end{equation}
so that 
\begin{equation}
  A^{\hat {i}}_{\ i}(t+\delta t)\approx(\delta^{\hat{i}}_{\ \hat{j}}-\delta t
  \,\epsilon^{\hat{i}}_{\ \hat{j}\hat{k}}H_{\hat{0}}^{\ \hat{k}})A^{\hat{j}}_{\ i}\,,
\end{equation}
where $\delta t$ represent an infinitesimal displacement along the time
coordinate. This is an infinitesimal rotation of the spatial part of the tetrad
basis about the angular velocity vector $\bfs{H}_{\hat{0}}$.

We will now obtain explicit algebraic expressions in terms of the 3-vector
fields $\bfs{H}_{\hat{\alpha}}$, $\bfs{D}_{\hat{\alpha}}$, $\bfs{E}^{\hat{\alpha}}$, 
$\bfs{B}^{\hat{\alpha}}$ for the projections of
the Sparling form $\rho_{\hat{\alpha}}$ and $s^j_{\ \hat{\alpha}}$, of which the latter are needed 
for
evolution. Expressing equation \eqref{eq:sparling-form} in component form in the
orthonormal frame and using the definition in equation
\eqref{eq:tmunu_definition}, we obtain
\begin{equation}
  \label{eq:ti_components}
  t^{\hat{\gamma}}_{\ \hat{\alpha}} = \half \left(\omega^{\hat{\sigma}}_{\ \hat{\alpha}
    \hat{\mu}} \omega^{\hat{\rho} \hat{\xi}}_{\ \ \hat{\nu}}
  \delta^{\hat{\tau}}_{\ \hat{\sigma}} + \omega^{\hat{\rho}}_{\ \hat{\sigma}
    \hat{\mu}} \omega^{\hat{\sigma} \hat{\xi}}_{\ \ \hat{\nu}}
  \delta^{\hat{\tau}}_{\ \hat{\alpha}} \right)
  \delta^{\hat{\nu}\hat{\mu}\hat{\gamma}}_{\hat{\tau}\hat{\rho}\hat{\xi}}\,,
\end{equation}
where we have made use of the generalized Kronecker delta to keep the notation
compact\footnote{The generalized Kronecker delta
$\delta_{\nu_{1}\dots\nu_{p}}^{\mu_{1}\dots\mu_{p}}$ is defined so that it
equals:
\begin{equation*}
  \begin{cases}
    +1&\text{ if }\nu_1\dots\nu_p
    \text{ are an even permutation of }\mu_1\dots\mu_p\\
    -1&\text{ if }\nu_1\dots\nu_p
    \text{ are an odd permutation of }\mu_1\dots\mu_p\\
    0&\text{ otherwise}\,.
  \end{cases}
\end{equation*}}.
Using the the relations given by equation \eqref{eq:Ricci_connection},
for a connection that is anti-symmetric with respect to its first two
indices equation \eqref{eq:ti_components} can be re-written as
\begin{equation}
  t^{\hat{\gamma}}_{\ \hat{\alpha}} = F^{\hat{\delta}}_{\ \hat{\beta}\hat{\alpha}}
  \star u_{\hat{\delta}}^{\ \hat{\beta}\hat{\gamma}}
  -\frac{1}{4} \delta^{\hat{\gamma}}_{\ \hat{\alpha}}
  F^{\hat{\delta}}_{\ \hat{\beta}\hat{d}}
  \star u_{\hat{\delta}}^{\ \hat{\beta}\hat{d}} \,,
\end{equation}
and taking the projections defined in equations \eqref{eq:proj_energy-momentum},
we obtain
\begin{align}
  \label{eq:rho0_fromEB}
  \rho_{\hat{0}} &= -\half \left(E^{\hat{\alpha}\hat{k}} D_{\hat{\alpha}\hat{k}}
  + B^{\hat{\alpha}\hat{k}} H_{\hat{\alpha}\hat{k}} \right) \\
  \label{eq:rhoi_fromEB}                             
  \rho_{\hat{i}} &= -\epsilon_{\hat{i}\hat{j}\hat{k}}
  B^{\hat{\alpha}\hat{j}} D_{\hat{\alpha}}^{\ \hat{k}} \\
  \label{eq:si0_fromEB}                               
  s^{\hat{i}}_{\ \hat{0}} &= -\epsilon^{\hat{i}\hat{j}\hat{k}}
  E^{\hat{\alpha}}_{\ \hat{j}} H_{\hat{\alpha}\hat{k}} \\
  \label{eq:sji_fromEB}                               
  s^{\hat{i}}_{\ \hat{j}} &= E^{\hat{\alpha}}_{\ \hat{j}} D_{\hat{\alpha}}^{\ \hat{i}}
  + B^{\hat{\alpha}\hat{i}} H_{\hat{\alpha}\hat{j}}\nonumber\\
  &-\half\delta^{\hat{i}}_{\ \hat{j}} \left(E^{\hat{\alpha}\hat{k}}D_{\hat{\alpha}\hat{k}}
  + B^{\hat{\alpha}\hat{k}} H_{\hat{\alpha}\hat{k}} \right)\,.
\end{align}

Although equations \eqref{eq:rho0_fromEB} and \eqref{eq:rhoi_fromEB} express
algebraic constraints between variables that are evolved with their own
differential equation, if the
momentum densities are evolved using a finite volume scheme, these relations
between the numerical representation of the variables should not be expected to
hold strictly. The reason is that the representation of the momentum densities is
that of a volume average, which does not need to coincide with the value of the
right hand side of the equations calculated at a given point (or with values
interpolated from a set of given points). However, these expressions may still
be useful to obtain additional information on these quantities, \eg to improve
interpolations. In this case the scheme would sacrifice the exact fulfillment of
these expressions in favor of machine precision conservation of energy and
momentum.

The last quantity for which we need to give an explicit expression is the
``force'' term given by equation \eqref{eq:Q}. After substituting
\eqref{eq:tmunu_definition} in \eqref{eq:Q} and decomposing $\bfs{F}_{\hat{\alpha}}$
as in equation \eqref{eq:F_components}, we obtain
\begin{align}
  \label{eq:Q0EB}
  Q_{\hat{0}}&=E^{\hat{i}\hat{j}} S_{\hat{i}\hat{j}} -
  E^{\hat{0}\hat{j}}P_{\hat{j}} \\
  \label{eq:QiEB}
  Q_{\hat{i}} 
  &= E^{\hat{k}}_{\ \hat{i}} S_{\hat{k} \hat{0}} + E^{\hat{0}}_{\ \hat{i}}
  P_{\hat{0}} + \epsilon_{\hat{i}\hat{j}\hat{k}} B^{\hat{l}\hat{j}}
  S^{\hat{k}}_{\ \hat{l}}
\end{align}

In order to close the system completely, it is necessary to specify a set of
relations between the non-gravitational energy and momentum $P_{\hat{\alpha}}$ and
their associated fluxes $S^{k}_{\ \hat{\alpha}}$, which will depend on the kind of
non-gravitational fields considered (\eg ideal fluid, electromagnetic fields, or
a scalar field).

\section{The \dgrem formulation}
\label{sec:full-system}

Finally, we can summarize here the equations obtained in the previous section in
order to describe the system completely. For each equation we indicate its
common name (or that of the equations more closely related to it) and the number
that labels it in the part of the text where it is discussed.

\subsubsection{Evolution equations}

\begin{align}
  \intertext{\bf{First Cartan structure equations}} 
  & \partial_t A^{\hat{i}}_{\ i} - \partial_i \beta^{\hat{i}} =
  -\alpha E^{\hat{i}}_{\ i} + \epsilon_{ilk} \beta^l B^{\hat{i} k}
  \tag{\ref{eq:Aupdate}} \\
  \intertext{\bf{First Bianchi identities}}
  & \partial_t \sqrt{\gamma} B^{\hat{\alpha} k}\nonumber\\
  &+\partial_i \sqrt{\gamma} (\alpha \epsilon^{ijk} E^{\hat{\alpha}}_{\ j}-\beta^i
  B^{\hat{\alpha} k} + \beta^k B^{\hat{\alpha} i}) 
  = 0 \tag{\ref{eq:BianchiFaraday}}\\
  \intertext{\bf{Einstein evolution equations}}
  & \partial_t \sqrt{\gamma} D_{\hat{\alpha}}^{\ k} - \partial_i \sqrt{\gamma}
  (\alpha \epsilon^{kij} H_{\hat{\alpha}j} + \beta^i D_{\hat{\alpha}}^{\ k} -
  \beta^k D_{\hat{\alpha}}^{\ i}) =\nonumber\\
  &-\sqrt{\gamma}(j^k_{\ a} + \kappa J^k_{\ a})
  \tag{\ref{eq:Einstein_evolD}}\\
  \intertext{\bf{Conservation of gravitational energy-momentum}}
  & \partial_t \sqrt{\gamma} \rho_{\hat{\alpha}} + \partial_i \sqrt{\gamma}
  j^i_{\ \hat{\alpha}}
  = -\kappa \sqrt{\gamma} Q_{\hat{\alpha}}
  \tag{\ref{eq:grav-energy-momentum-cons}}  \\
  \intertext{\bf{Conservation of `matter' energy-momentum}}
  & \partial_t \sqrt{\gamma} P_{\hat{\alpha}} + \partial_i \sqrt{\gamma}
  J^k_{\ \hat{\alpha}} = \ \ \sqrt{\gamma} Q_{\hat{\alpha}}
  \tag{\ref{eq:matt-energy-momentum-cons}} \\
  \intertext{\bf{Auxiliary evolution equation for $\sqrt{\gamma}$}}
  \tag{\ref{eq:sqrtgammaevol}}
  & \partial_t \sqrt{\gamma} -\partial_i \sqrt{\gamma} \beta^i
  = \frac{5}{4}\sqrt{\gamma} D_{\hat{k}}^{\ \hat{k}}
\end{align}
%

\subsubsection{Differential constraints}

\begin{align}
  \intertext{\bf{First Cartan structure equations}}
  E^{\hat{0}}_{\ i} &= -\partial_i \ln \alpha
  \tag{\ref{eq:EBconstraints}.a}\\
  B^{\hat{0}i} &= 0 \tag{\ref{eq:EBconstraints}.b}\\
  B^{\hat{i}k} &= \epsilon^{ijk} \partial_i A^{\hat{i}}_j \tag{\ref{eq:EBconstraints}.c}\\
  \intertext{\bf{First Bianchi identities}}
  \partial_i \sqrt{\gamma} B^{\hat{\alpha} i} &= 0 \tag{\ref{eq:BianchiDivB}}\\
  \intertext{\bf{Hamiltonian and momentum constraints}}
  \partial_i \sqrt{\gamma} D_{\hat{\alpha}}^{\ i}
  &= \sqrt{\gamma} (\rho_{\hat{\alpha}} + \kappa P_{\hat{\alpha}})
  \tag{\ref{eq:Einstein_constraintsD}}
\end{align}

\subsubsection{Constitutive relations}

\begin{align}
  H_{\hat{k} \hat{i}} &= - B_{\hat{i} \hat{k}}
  + \half \delta_{\hat{k} \hat{i}} B_{\hat{l}}^{\ \hat{l}}
  - \epsilon_{\hat{k}\hat{i}\hat{l}} E_{\hat{0}}^{\ \hat{l}} \tag{\ref{eq:Hk_EB}} \\
  E^{\hat{j}}_{\ \hat{k}} &= - D_{\hat{k}}^{\ \hat{j}} -\half
  \delta^{\hat{j}}_{\ \hat{k}} D_{\hat{l}}^{\ \hat{l}}
  +\epsilon^{\hat{j}}_{\ \hat{k}\hat{l}} H_{\hat{0}}^{\ \hat{l}}
  \tag{\ref{eq:Ek_DH}}\\
  \intertext{\bf{Gravitational energy-momentum current}}
  j^k_{\ \hat{\alpha}} &= \alpha s^{k}_{\ \hat{\alpha}} - \beta^k\rho_{\hat{\alpha}}
  \tag{\ref{eq:j_gravitational}}\\
  s^{\hat{i}}_{\ \hat{0}} &= -\epsilon^{\hat{i}\hat{j}\hat{k}}
  E^{\hat{\alpha}}_{\ \hat{j}} H_{\hat{\alpha}\hat{k}}
  \tag{\ref{eq:si0_fromEB}} \\
  s^{\hat{i}}_{\ \hat{j}} &= E^{\hat{\alpha}}_{\ \hat{j}} D_{\hat{\alpha}}^{\ \hat{i}}
  + B^{\hat{\alpha}\hat{i}} H_{\hat{\alpha}\hat{j}}\nonumber\\
  -&\half \delta^{\hat{i}}_{\ \hat{j}}\left(E^{\hat{\alpha}\hat{k}}D_{\hat{\alpha}\hat{k}}
  + B^{\hat{\alpha}\hat{k}} H_{\hat{\alpha}\hat{k}} \right)
  \tag{\ref{eq:sji_fromEB}}\\
  \intertext{\bf{`Matter' energy-momentum current}}
  J^k_{\ \hat{\alpha}} &= \alpha S^{k}_{\ \hat{\alpha}} - \beta^k P_{\hat{\alpha}}
  \tag{\ref{eq:J_matter}}\\
  \intertext{\bf{`Gravitational force'}}
  Q_{\hat{0}}&=E^{\hat{i}\hat{j}} S_{\hat{i}\hat{j}} -
  E^{\hat{0}\hat{j}}P_{\hat{j}} \tag{\ref{eq:Q0EB}} \\
  Q_{\hat{i}} 
  &= E^{\hat{k}}_{\ \hat{i}} S_{\hat{k} \hat{0}} + E^{\hat{0}}_{\ \hat{i}}
  P_{\hat{0}} + \epsilon_{\hat{i}\hat{j}\hat{k}} B^{\hat{l}\hat{j}}
  S^{\hat{k}}_{\ \hat{l}}
  \tag{\ref{eq:QiEB}}
\end{align}

\subsubsection{Algebraic constraints}

\begin{align}
  D_{\hat{0}}^{\ \hat{i}} &= -\epsilon^{\hat{i}\hat{j}\hat{k}} B_{\hat{j} \hat{k}}
  \tag{\ref{eq:Dk_EB}} \\
  D_{\hat{i}\hat{j}} &= D_{\hat{j}\hat{i}} \tag{\ref{eq:B0_DH}} \\
  \rho_{\hat{0}} &= -\half \left(E^{\hat{\alpha}\hat{k}} D_{\hat{\alpha}\hat{k}}
+ B^{\hat{\alpha}\hat{k}} H_{\hat{\alpha}\hat{k}} \right) \tag{\ref{eq:rho0_fromEB}}\\
  \rho_{\hat{i}} &= -\epsilon_{\hat{i}\hat{j}\hat{k}}
B^{\hat{\alpha}\hat{j}} D_{\hat{\alpha}}^{\ \hat{k}} \tag{\ref{eq:rhoi_fromEB}}
\end{align}

\subsubsection{Free quantities}

The fields $\alpha$, $\beta^i$ and $H_{\hat{0}}^{\ k}$ are not determined by any
equation and can be chosen arbitrarily. The matter energy-momentum fluxes
$S^k_{\ \hat{\alpha}}$ are not determined by any of the equations here, but
depend on the specific properties of the matter fields.

\subsection{Properties of the formulation}

The final system of equations is in a form that closely resembles those of
electromagnetism in the 3+1 decomposition, with the difference that the
gravitational field is represented not by one, but by four
``electromagnetic-like'' fields $(E^{\hat{\alpha}}_{\ i},B^{\hat{\alpha} i})$,
and that due to the particular choice of the observers frame the field corresponding to 
$\hat{\alpha}=0$ is purely ``electric''.

Being more explicit in this analogy, the gauge variables $\alpha$
and $\beta^{\hat{i}}$,
or more specifically the quantities $-\beta^{\hat{i}}$ and $\ln
\alpha$, play a role analogous to that of the scalar potential in
electromagnetism; while the
components of the spatial part of the tetrad play the role of the vector
potential, as can be seen from equations \eqref{eq:Aupdate} and
\eqref{eq:EBconstraints}.

The first Bianchi identities take a form analogous to that of the Faraday
equation \eqref{eq:BianchiFaraday} and the Gauss law for magnetism
\eqref{eq:BianchiDivB}, while the Einstein equations take that of the
Amp\`ere-Maxwell equation \eqref{eq:Einstein_evolD} and the Gauss law for
electricity \eqref{eq:Einstein_constraintsD}, with the sum of matter and
gravitational energy-momentum playing the role of the electric current, which
satisfies an exact conservation law (see
Eqs.~\eqref{eq:grav-energy-momentum-cons}, \eqref{eq:matt-energy-momentum-cons}
and \eqref{eq:rhoP_tot}).

Although not of immediate use for a numerical implementation, it is interesting
to notice other similarities of the equations with those of electromagnetism.
For instance, the expressions for the gravitational energy-momentum density and
fluxes are analogous to those given by Minkowski's energy-momentum tensor for
the electromagnetic field in material media \citep{Jackson:1998},
and contain an expression related to the transport of gravitational
energy \eqref{eq:si0_fromEB} that is analogous to the Poynting vector in electrodynamics.
The force terms that
describe the exchange between matter and the gravitational field in
Eqs.~\eqref{eq:grav-energy-momentum-cons} and
\eqref{eq:matt-energy-momentum-cons} have a form similar to that of the work
done by the electric field on a system of charges \eqref{eq:Q0EB} and to the
Lorentz force \eqref{eq:QiEB}.

However, there are also important differences with respect to Maxwell's
equations. The most noticeable one is that the inhomogeneous equations contain
source terms quadratic in the fields, which represent the fact that the
gravitational energy-momentum current $j^{\mu}_{\ \hat{\alpha}}$ is itself a
source for the gravitational field $D^{\ \mu}_{\hat{\alpha}}$. Another important
difference is that the presence of the square root of the metric determinant
$\sqrt{\gamma}=\det (A^{\hat{\alpha}}_{\ \mu})$ eliminates the gauge freedom
that in electrodynamics allows one to replace $A^{\prime}_{\mu}\rightarrow
A_{\mu}+\partial_\mu\psi$, where $\psi$ is a scalar function and $A_\mu$ the
vector potential. This prevents us from choosing to solve the ``Faraday
equation'' \eqref{eq:BianchiFaraday} in place of the evolution equation for the
vector potential \eqref{eq:Aupdate} and forces us to solve the latter in order
to know the transformation coefficients from the
``laboratory frame'' to the tetrad frame where the constitutive relations
\eqref{eq:Hk_EB} and \eqref{eq:Ek_DH} are valid.

Although the gauge freedom of electrodynamics does not exist for this system, it
posses other gauge freedoms. These come in through the quantities for which
neither the Cartan structure equations nor the Einstein equations provide an
evolution equation, namely the components of the vector normal to the
hypersurface $n^{\mu}=(1/\alpha,-\beta^i/\alpha)$ and the ``magnetic field''
$H_{\hat{0}\mu}$. While the freedom in choosing $n^{\mu}$ represents the freedom
to foliate the spacetime in different sets of 3D hypersurfaces and to perform
spatial translations of the lines of constant spatial coordinates, the freedom
to choose $H_{\hat{0}\mu}$ represents the liberty to perform rotations of the
spatial part of the tetrad from one slice to the other (see
\secref{closing_system}). Although in contrast to electromagnetism these gauge
freedoms do not leave unchanged the vector fields $E^{\hat{\alpha}}_{\ i},
B^{\hat{\alpha}i}, D_{\hat{\alpha}}^{\ i}, H_{\hat{\alpha} i}$, the Einstein
tensor at a given point, given by Eq.~\eqref{eq:einstein_from_U_t} will be the
same object regardless of the foliation and the orientation of the basis
vectors. Going beyond GR to include torsion, the system does
contain an additional freedom that leaves the fields
unchanged\footnote{This freedom comes from
	regarding the field strength $\bfs{F}^{\hat{\alpha}}$
	as the sum of the torsion $\bfs{\Xi}^{\hat{\alpha}}$
	and the product $\cform^{\hat{\alpha}}_{\ \hat{\beta}}\wedge\base^{\hat{\beta}}$
    (\cf equations \ref{eq:BianchiI} and \ref{eq:F_definition}).
    For a set of 2-forms $\bfs{S}_{\hat{\alpha}}(\bfs{F}^{\hat{\alpha}})$
    which has the same functional dependence on $\bfs{F}^{\hat{\alpha}}$
    as that of $\star\bfs{u}_{\hat{\alpha}}(\bfs{F}^{\hat{\alpha}})$
    in GR, the Lagrangian
    \begin{equation}
        \nonumber
    	L[A^{\hat{\alpha}}_{\ \mu},\partial_{\lambda}A^{\hat{\alpha}}_{\ \mu}]
    	=\frac{\sqrt{-g}}{4\kappa} F^{\alpha}_{\ \hat{\mu}\hat{\nu}}
    	S_{\alpha}^{\ \hat{\mu}\hat{\nu}}
    \end{equation}
will lead to equations of motion identical to those presented here regardless of
the amount of torsion contained in $\bfs{F}^{\hat{\alpha}}$. For
$\bfs{\Xi}^{\hat{\alpha}}=0$, this Lagrangian is equivalent to the
Einstein-Hilbert Lagrangian up to a boundary term, and for the extreme case
$\bfs{F}^{\hat{\alpha}}=\bfs{\Xi}^{\hat{\alpha}}$ it correspond to that of the
teleparallel equivalent of GR, with $\bfs{S}_{\hat{\alpha}}$ identified as the
superpotential \citep[\cf Appendix C of ][]{AldrovandiPereira2013}}. It is
conceivable that, similarly to the gauge variables $\alpha$ and $\beta^i$, the
vector $\bfs{H}_{\hat{0}}$ could play an important role in the numerical
stability of the system, and more studies on a proper way to handle this
additional freedom are required.

Related to its similarity to the Maxwell equations, the \dgrem system also
posses the important properties of being first order in spatial and temporal
derivatives, and being expressible as a system of flux-balanced laws. As
mentioned in the Introduction such properties make possible the use of the huge
amount of technology developed to simulate such systems.

Finally, being formulated as a system of equations in differential forms and
exterior derivatives, it is possible to retrieve a natural constraint-preserving
discretization, which would also make redundant some of the evolution equations,
reducing the number of variables needed for evolution. An example of such
discretization with a reduced number of variables will be presented in the next
Section.

\subsection{A geometric interpretation}
\label{sec:GR-discretization}

One of the advantages of using a constrained transport scheme is that many of
the equations in the system described in \secref{full-system} become redundant
when using the proper discretization. The reason is that
if a consistent discretization is adopted for all the equations,
those, that are exterior derivatives of others are
automatically fulfilled. In particular, the scheme described here
requires only the evolution of Eqs.~\eqref{eq:Aupdate} and
\eqref{eq:Einstein_evolD} to satisfy all equations in the
system summarized in Section \ref{sec:full-system}.
The equations presented in this Section
are only those related to the evolution of spacetime, while
the matter sector is assumed to be evolved with an unspecified
scheme that is conservative for energy-momentum.

Similarly as done for the wave equation in \secref{ext_systems},
we will obtain a constraint-preserving discretization
on the hypersurface $\Sigma_t$
by first applying Cartan's `magic' formula \eqref{eq:cartan_magic},
followed by integrating the differential forms on their respective
sub-manifolds and applying Stoke's theorem \eqref{eq:stokes}.

The first step of the procedure yields the equations
\begin{align}
  \mathcal{L}_{\bfs{e}_t}\base^{\hat{\alpha}}-\ext(\bfs{e}_t\cdot\base^{\hat{\alpha}})
  &=\bfs{e}_t\cdot\bfs{F}^{\hat{\alpha}}\,, \\
  \mathcal{L}_{\bfs{e}_t}\bfs{u}_{\hat{\alpha}}-\ext(\bfs{e}_t\cdot\bfs{u}_{\hat{\alpha}})
  &=\bfs{e}_t\cdot(\bfs{t}_{\hat{\alpha}}+\kappa\bfs{T}_{\hat{\alpha}})\,.
\end{align}
which can also be written as
\begin{align}
  \label{eq:A_evol_forms}
  \partial_t \mathcal{A}^{\hat{i}}-\ext\beta^{\hat{i}}&=\mathcal{E}^{\hat{i}}\,,\\
  \label{eq:D_evol_forms}
  \partial_t\mathcal{D}_{\hat{\alpha}}-\ext\mathcal{H}_{\hat{\alpha}}
  &=\mathcal{J}_{\hat{\alpha}}\,,
\end{align}
where
\begin{align}
  \mathcal{A}^{\hat{i}} &= A^{\hat{i}}_{\ i} \ext x^i \,, \\
  \mathcal{E}^{\hat{i}} &= (-\alpha E^{\hat{i}}_{\ i} +
  \varepsilon_{ilk} \beta^l B^{\hat{i} k}) \ext x^i \,, \\
  \mathcal{H}_{\hat{\alpha}} &= (\alpha H_{\hat{\alpha} i} +
  \varepsilon_{ilk} \beta^l D_{\hat{\alpha}}^{\ k}) \ext x^i \,,
\end{align}
and
\begin{align}
  \mathcal{D}_{\hat{\alpha}} &=\varepsilon_{ijk} D_{\hat{\alpha}}^{\ i} 
  \left(\half \ext x^j \wedge \ext x^k\right) \,, \\
  \mathcal{J}_{\hat{\alpha}} &=\varepsilon_{ijk}
  (t^{i}_{\ \hat{\alpha}} + \kappa T^{i}_{\ \hat{\alpha}})
  \left(\half \ext x^j \wedge \ext x^k\right) \,,
\end{align}
and where the system is closed by the constitutive relations
and by adopting a consistent discretization for the forms
\begin{equation}
  \mathcal{B}^{\hat{\alpha}} =\varepsilon_{ijk} B^{\hat{\alpha} i} 
\left(\half \ext x^j \wedge \ext x^k\right) \,, \\
\end{equation}
in order to obtain $B^{\hat{\alpha} i}$ from equation \eqref{eq:EBconstraints}.

\begin{figure*}
	\centering
	\includegraphics[width=0.9\linewidth]{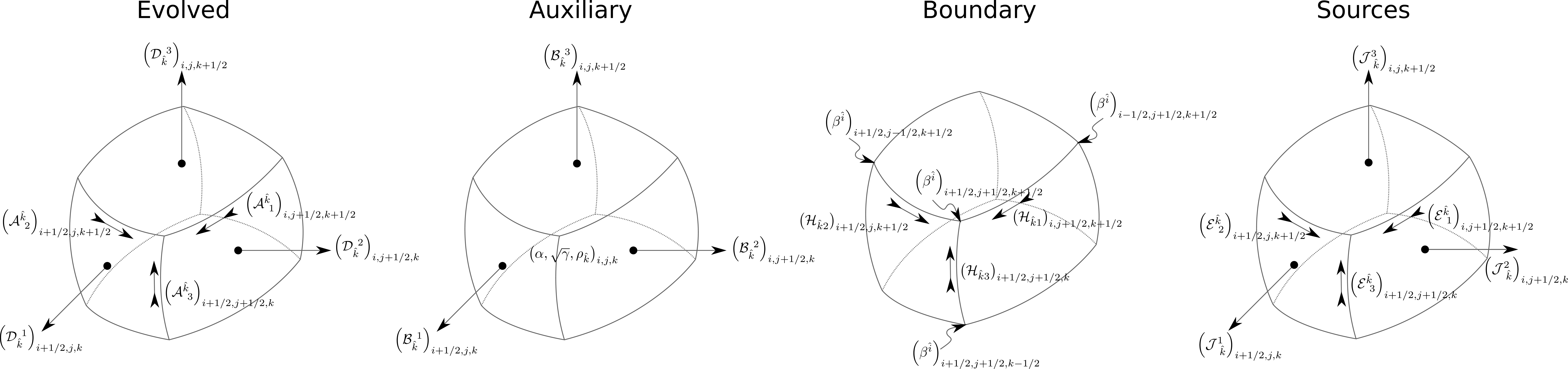}
	\caption{Collocation of variables for a constraint-preserving discretization.
	These are classified in four categories: `evolved' variables are those
    obtained by integrating the evolution equations of the scheme,
    `boundary' variables are those localized at the boundaries of the
    regions where the evolved variables are defined, and `source' variables
    are those sharing the same spatial location as the evolved variables.
    Finally, `auxiliary' variables are those that can be obtained from the
    evolved variables, but are neither sharing their spatial location
    nor that of their boundaries.}
	\label{fig:ct-scheme}
\end{figure*}

Each term in equation \eqref{eq:A_evol_forms} (\eqref{eq:D_evol_forms}) is a
1-form (a 2-form in), and thus an integrand over a 1D (2D) submanifold. 
We then choose
to integrate them over zone edges and zone faces, respectively.
After applying Stokes' theorem and replacing exterior derivatives with
evaluations of forms at zone vertices and zone edges, the resulting discretization
is as shown in Figure\,\ref{fig:ct-scheme}, and
in principle could be able to preserve to machine accuracy simultaneously
the Bianchi identities \eqref{eq:BianchiDivB},
the Einstein constraints \eqref{eq:Einstein_constraintsD},
as well as the global conservation of the sum of gravitational
plus matter energy-momentum \eqref{eq:rhoP_tot},
provided that they are satisfied in the initial data, by the
mechanism described in Section \ref{sec:ext_systems}.

\section{Conclusions}
\label{sec:conclusions}

By expressing the equations that govern space-time dynamics in general
relativity in the language of exterior calculus and projecting them onto
3-dimensional space-like hypersurfaces, we have obtained a new 3+1 formulation
of the field equations of general relativity. This new formulation, which we name
\dgrem, shows a surprising resemblance to the equations of relativistic \ac{MHD}
and to \ac{EM} in material media. The system, summarized in
\secref{full-system}, consists of a set of first-order evolution equations, in
conservative form, and a set of algebraic, divergence and curl constraints,
closed by a set of constitutive relations.

The similarities with 3+1 electrodynamics make explicit some important features
of general relativity, such as the global conservation of total energy-momentum
currents (in analogy to that of electric current), the fact that both the
gravitational and matter energy momentum act as sources of the gravitational
field, as well as the energy-momentum exchange between the gravitational and
matter sectors.

Additionally, the \dgrem formulation exhibits several interesting properties
from the point of view of numerical implementations. Being first order and
flux-conservative, it is suitable for the application of high-resolution
shock-capturing schemes such as finite-volume and finite-element methods. In
particular the formulation contains a global conservation equation for the sum
of gravitational and ``matter'' energy-momentum in which source terms have been
eliminated, and which opens the possibility of applying techniques such as
first-order flux limiting to ensure positivity of energy-momentum densities.

As shown in \secref{GR-discretization}, the expression of the formulation as a
set of equations in differential forms permits to integrate them over mesh zones
and use Stoke's theorem to obtain a natural staggered discretization potentially
suitable for machine-precision constraint-preserving schemes. One such scheme
could potentially reduce the number of evolution variables to a minimum of 21,
both by not requiring extra variables to clean the constraints and by making
redundant some of the equations.

Although a staggered scheme would enforce at machine-precision both the
fulfillment of the Einstein constraints and the conservation of energy-momentum,
these advantages may be limited in practice for general relativistic
hydrodynamic simulations due to the adoption of a floor model as it is
customarily done to handle vacuum regions.

However, these techniques could in principle also be exploited in fully general
relativistic N-body simulations, which could recycle the infrastructure
developed for \ac{PIC} simulations of collisionless plasmas, in which both
staggered schemes and divergence cleaning techniques have been successfully
applied.

In the same way, it is conceivable that resemblance of the form taken by the
constraints of this formulation to Gauss' laws in electromagnetism could present
advantages for the computation of initial data by recycling techniques used to
solve the Poisson equation.

Finally, another benefit of deriving the system as a set of equations in terms
of differential forms and exterior derivatives is that they naturally give
relations between quantities evolved inside mesh cells and quantities evaluated
at cell boundaries, regardless of the shape of the cells. This makes them
particularly suitable for simulations using non-Cartesian coordinates and
unstructured meshes.

Finally, the matter sector of the Einstein field equations (including
relativistic dissipative fluid dynamics) can be also formulated in the language
of differential forms and exterior calculus \citep{PTRSA2020,Torsion2019}, and
thus can be relatively easily incorporated in the constrained transport
computational scheme discussed in \secref{GR-discretization}.

Together with the promising properties summarized above, there are still some
questions regarding \dgrem that need to be answered for a successful numerical
implementation. The most important one is perhaps on its hyperbolicity, and how
it could depend on gauge choices and on the new degrees of freedom given by
spatial rotations of the tetrads between different hypersurfaces.

Other particulars of an actual numerical implementation are still under
development, and will be part of a future work.

\section*{Acknowledgements}

During the development of this project, the authors became aware of a work in
preparation by I. Peshkov and E. Romenski on a first-order reduction of pure
tetrad teleparallel gravity which includes as a special case a system of
equations identical to that presented here, and which served as an independent
verification of our derivations. HO is grateful to M. DeLaurentis, B. Ripperda,
M. Moscibrodzka, O. Porth, A. Jim\'enez-Rosales, J. Vos, J. Davelaar, C. Brinkerink,
T. Bronzwaer and A. Cruz-Osorio for useful discussions on the formulation, and to
V. Ardachenko for sharing thoughts on the relation between constraint-preserving
discretizations and equations in exterior derivatives. 
ERM thanks F. Pretorius for useful discussions related to this work. 
HO acknowledges support from a Virtual Institute of Accretion (VIA)
postdoctoral fellowship from the Netherlands Research School for Astronomy
(NOVA). IP would like to thank the Italian Ministry of Instruction, University and Research (MIUR) 
to support his research with funds coming from PRIN Project 2017 No. 2017KKJP4X entitled 
"Innovative numerical methods for evolutionary partial differential equations and applications". 
ERM gratefully acknowledges support from postdoctoral fellowships at the
Princeton Center for Theoretical Science, the Princeton Gravity Initiative,
and the Institute for Advanced Study.

\appendix

\section{A small primer on differential forms and exterior calculus}
\label{sec:diff_forms}

We collect here some fundamental results about differential forms and exterior
calculus, necessary to follow the derivations in this work. The modern theory of
differential forms and exterior calculus stems from the work of \'Elie Cartan in
the first half of the twentieth century, and the literature regarding this field
is by now very extensive. For further reading we refer the reader to
\citep{Nakahara2018, Frauendiener2006, Burton2003} and references therein, which
are the sources this primer is based on. Note that we quote definitions and
results in the form they assume in the spacetime of \ac{GR}, \ie a 4-dimensional
Lorentzian manifold, indicated by the symbol $\mathcal{M}$. We refer the
interested reader to the literature for statements valid in more general
settings.

A sum of the form
\begin{equation}
  \bfs{F}=F_a\base^a
\end{equation}
is called a $1$-differential form, or simply a $1$-form, and $F_a$ are its
components. $1$-forms are therefore identical to covariant vectors. More
generally, $p$-differential forms (in the following simply $p$-forms) are
rank-$p$ totally antisymmetric covariant tensors on $\mathcal{M}$.
The differential forms of highest possible degree are $4$-forms, since for
higher degrees the antisymmetry requirement would make any differential form
vanish identically. $0$-forms are defined as scalar functions on $\mathcal{M}$
(scalar fields).

The set of $p$-forms at a point $P$ of $\mathcal{M}$ forms a
$\binom{4}{p}$-dimensional vector space. Therefore the dimensions of the spaces
of $0$-, $1$-, $2$-, $3$- and $4$-forms (and the number of components of any
form in one of these spaces) are respectively 1, 4, 6, 4, 1.

For the rest of this section, let $\bfs{A}$ and $\bfs{B}$ be generic $p$- and
$q-$forms respectively. We define an operation that acts on two such forms
to produce a $(p+q)$-form. This is referred to as the \textit{exterior product}
or \textit{wedge product}, and it is defined as:
\begin{equation}
  \bfs{A}\wedge\bfs{B} := \textnormal{Alt}(\bfs{A}\otimes\bfs{B})\,,
\end{equation}
where $\otimes$ is the standard tensor product and $\textnormal{Alt}(\bfs{T})$
denotes is the totally antisymmetric part of the tensor $\bfs{T}$.
The components of a the result of the wedge product are therefore:
\begin{align}
  &(\bfs{A}\wedge\bfs{B})_{a_1\dots a_{p+q}}=\nonumber\\
  &\frac{1}{(p+q)!}\sum_{P\in S}\sgn(P)A_{a_{P(1)}\dots a_{P(p)}}
  B_{b_{P(p+1)}\dots b_{P(p+q)}}\,,
\end{align}
where $S$ is the set of all possible permutations of $p+q$ elements, $P$ is one
such permutation and $\sgn(P)$ equals $+1$ for even permutations and $-1$ for
odd ones. Using a shorthand notation common in the \ac{GR} literature, this
formula can be written as:
\begin{equation}
  (\bfs{A}\wedge\bfs{B})_{c_1\dots a_{p+q}} = A_{[a_1\dots a_p}B_{a_{p+1}\dots a_{p+q}]}\,.
\end{equation}

The exterior product is associative, and more importantly it satisfies the
relation
\begin{equation}
  \bfs{A}\wedge\bfs{B} = (-1)^{pq}\bfs{B}\wedge\bfs{A}\,.
\end{equation}
This in particular implies that for $1$-forms the exterior product is
antisymmetric.

Recall that the set $\base^a$ is a basis of the vector space of
$1$-forms. Leveraging the antisymmetry of the exterior product for $1$-forms, it
can be seen that the set of elements of the form
\begin{equation}
  \base^{a_1}\wedge\dots\wedge\base^{a_p}\,,
\end{equation}
\ie the exterior product of $p$ elements of the basis of $1$-forms, constitutes
a basis for the vector space of $p$-forms. For example, a basis for the space of
$2$-forms in a 4-dimensional spacetime is
\begin{align}
  \{\base^0\wedge&\base^1,\base^0\wedge\base^2,\base^0\wedge\base^3,\nonumber\\
  &\base^1\wedge\base^2,\base^1\wedge\base^3,\base^2\wedge\base^3\}\,,
\end{align}
which as noted above has 6 elements.

A $1$-form defines a linear operator acting on vectors and producing a real
number, so that the result of a $1$-form $\bfs{F}$ acting on a vector $\bfs{X}$
can be written
\begin{equation}
  \bfs{F}(\bfs{X}) = F_aX^a = \langle\bfs{F},\bfs{X}\rangle\,,
\end{equation}
where the last equality shows that this is nothing but the interior product
between vectors and their duals induced by the metric.

The \textit{interior product} is instead an operation between a $p$-form and a
vector $\bfs{X}$, which gives as result a $(p-1)$-form according to the
definition:
\begin{equation}
  (\iota_{\bfs{X}}\bfs{A})_{a_2\dots a_p} := X^{a_1}A_{a_1a_2\dots a_p}\,.
\end{equation}
While the inner product and the interior product should not be confused, the
latter is in a sense an extension of the former, since
$\iota_{\bfs{X}}\bfs{F}=\langle\bfs{X},\bfs{F}\rangle=\bfs{F}(\bfs{X})$.

As stated above, $p$-forms are antisymmetric $(0,p)$-tensors, and as tensors
they are acted upon by the standard partial and covariant derivatives. There is
however another type of derivation which affects these objects (and is instead
not defined for more general tensors). This is called the \textit{exterior
  derivative}, and denoted by the symbol $\ext$. It can be defined by stating
that the exterior derivative of a form $\bfs{A}=A_{a_1\dots
  a_p}\base^{a_1}\wedge\dots\wedge\base^{a_p}$ is
\begin{align}
  \ext\bfs{A}=
  (\partial_bA_{a_1\dots
    a_p})\base^b\wedge\base^{a_1}\wedge\dots\wedge\base^{a_p}\,.
  \label{eq:ext-derivatives}
\end{align}
Since the exterior products automatically antisymmetrize the coefficients, this
definition implies that the components of the result can be written as
\begin{align}
  (\ext\bfs{A})_{ba_1\dots a_p}=\partial_{[b}A_{a_1\dots a_p]}\,.
  \label{eq:ext-derivatives}
\end{align}

The exterior derivative associates to any $p$-form a $(p+1)$-form, and it
clearly does not depend on the metric or on any other additional structure on
the manifold. Despite the partial derivative being used in its definition, the
components of the exterior derivative form the components of a tensor, \ie
objects obtained by applying it transform as tensors under changes of basis.

Note that as the partial derivative, the exterior derivative is a linear
operation, however it exhibits a modified Leibniz rule with respect to the
exterior product:
\begin{equation}
  \label{eq:leibniz_ext}
  \ext(\bfs{A}\wedge\bfs{B})=
  \ext\bfs{A}\wedge\bfs{B}+(-1)^p\bfs{A}\wedge\ext\bfs{B}\,.
\end{equation}

Another fundamental property of the exterior derivative, which is leveraged at
several points in the present work, is its nilpotency:
\begin{equation}
  \ext\ext\bfs{A} = 0\,.
\label{eq:nilpotent}
\end{equation}

Note that having defined the exterior derivative and interior product, the
definition of the Lie derivative of a $p$-form $\bfs{A}$ along a vector
$\bfs{X}$ becomes particularly compact and easy to recall:
\begin{equation}
  \label{eq:cartan_magic}
  \mathcal{L}_{\bfs{X}}\bfs{A}=\ext\iota_{\bfs{X}}\bfs{A} +
  \iota_{\bfs{X}}\ext\bfs{A}\,.
\end{equation}
This is known as ``Cartan's magic formula''.

There also exists a definition of a \textit{exterior covariant derivative}, but
to state it we need to first introduce so-called tensor-valued differential
forms. So far in this section we only have used real-valued differential forms,
\ie form that when acting upon (sets of) vectors return a real value. However in
the main text we make extensive use of tensor-valued forms, which return a
collection of real values instead. These forms can be seen as collections of
real-valued forms, each member of the collection labeled by indices. Such an
object are the \textit{connection forms} $\cform^a_{\ b}$, a collection of
$1$-forms, defined by
\begin{equation}
  \nabla_{\bfs{e}_a}\bfs{e}_b=\cform^c_{\ b}(\bfs{e}_a)\bfs{e}_c\,.
  \label{eq:connection_forms}
\end{equation}
If the connection is chosen as the usual Levi-Civita connection, then
$\cform^\mu_{\ nu}=\Gamma^\mu_{\ \lambda\nu}\base^\lambda$ where
$\Gamma^\mu_{\ \lambda\nu}$ are the usual Christoffel symbols. In general
however the connection forms encode any arbitrary connection.

A few comments are in order. First of all, despite the possibly confusing
notation, note that $\cform^a_{\ b}$ is \textit{not} a rank-2 tensor of type
$(1,1)$. It is collection of $1$-forms, which becomes apparent by noting that it
is defined as the product of the basis $1$-forms and a collection of numbers.
Secondly, just as the components of the Christoffel symbols do not transform as
the components of a tensor, neither do the components of the object that the
connection forms yield when applied to a vector. In this sense the name
``tensor-valued form'' if applied to the connection forms is a misnomer, since
the components of the object yielded by such a form do not, in general,
transform as a tensor. The locution ``collection of $p$-forms'' while possibly
less descriptive, is also more appropriate. In light of this, we refer to the
indices of the connection $1$-forms in \eqref{eq:connection_forms} as
``non-tensorial'' indices. In the main text we deal with collection of forms,
some of which are non-tensorial like the connection forms and others instead are
proper tensor-valued forms, \ie their components do transform as those of
tensors.

The connection $1$-forms allow us to finally define the \textit{exterior
  covariant derivative} of a tensor-valued $p$-form by
\begin{align}
  &\covext\bfs{T}^{a\dots d}_{\hphantom{a\dots d}e\dots h}=
  \ext\bfs{T}^{a\dots d}_{\hphantom{a\dots d}e\dots h}+\nonumber\\
  &+\cform^a_{\ i}\wedge\bfs{T}^{i\dots d}_{\hphantom{a\dots d}e\dots h}
  +\cform^d_{\ i}\wedge\bfs{T}^{a\dots i}_{\hphantom{a\dots d}e\dots
    h}+\nonumber\\
  &-\cform^i_{\ e}\wedge\bfs{T}^{a\dots d}_{\hphantom{a\dots d}i\dots h}
  -\cform^i_{\ h}\wedge\bfs{T}^{a\dots d}_{\hphantom{a\dots d}e\dots i}\,.
\end{align}
Note however that this operation is only defined when applied on a form that is
tensor-valued in the strict sense, \ie when its indices are actually tensorial
and transform as the components of a tensor. Under this condition, the indices
of the result of applying the covariant exterior derivative will also transform
as those of a tensor.

In what follows we go back to real-valued forms. As a consequence of the
antisymmetry of differential forms, all the $4$-forms (\ie the highest possible
degree forms in a $4$-dimensional manifold) are multiples of a single $4$-form,
called \textit{volume form} or \textit{metric volume element}, and defined as
\begin{equation}
  \bfs{\varepsilon}=\sqrt{-g}\,\base^0\wedge\base^1\wedge\base^2\wedge\base^3\,.
\end{equation}
Its components can be written as
\begin{equation}
  \varepsilon_{abcd}=\sqrt{-g}\epsilon_{abcd}\,,
\end{equation}
where as anticipated in \secref{notation}, $g$ is the determinant of the metric
and the Levi-Civita symbol $\epsilon_{abcd}$ equals $+1$ or $-1$ depending on
whether $(a,b,c,d)$ is an even or an odd permutation of $(0,1,2,3)$. Note also
that raising the components of the volume element with the metric results in
\begin{equation}
  \varepsilon^{abcd}=-\frac{1}{\sqrt{-g}}\epsilon^{abcd}\,.
\end{equation}

It is also useful to note these properties of the volume form and Levi-Civita
symbol when restricted to purely spatial, tridimensional hypersurfaces, which
are used extensively in the main text:
\begin{align}
  \varepsilon_{0ijk}&=-\alpha\varepsilon_{ijk} \\
  \varepsilon^{0ijk}&=\frac{1}{\alpha}\varepsilon^{ijk} \\
  \varepsilon_{ijk}&=\sqrt{\gamma}\epsilon_{ijk}\\
  \varepsilon^{ijk}&=\frac{1}{\sqrt{\gamma}}\epsilon^{ijk}\,.
\end{align}
Furthermore, we note that in a non-coordinate, orthonormal frame, $g=-1$ so that
\begin{equation}
  \varepsilon_{\hat{\alpha}\hat{\beta}\hat{\gamma}\hat{\delta}}=
  \epsilon_{\hat{\alpha}\hat{\beta}\hat{\gamma}\hat{\delta}}
  \quad\textnormal{and}\quad
  \varepsilon_{\hat{i}\hat{j}\hat{k}}=
  \epsilon_{\hat{i}\hat{j}\hat{k}}\,.
\end{equation}
Accordingly in such a case (but not in general) we can write the former for the
latter and vice versa.

As outlined above the vector space of $p$-forms and that of $(4-p)$-forms have
the same dimension. Therefore it is possible to build an isomorphism between
these spaces. A very important such isomorphism is the \textit{Hodge duality},
represented by the symbol $\star$. The components of the Hodge dual can be
obtained as
\begin{equation}
  (\bfs{\star A})_{a_{p+1}\dots a_4} =
  \varepsilon_{a_1\dots a_p,\,a_{p+1}\dots a_4} A^{a_1\dots a_p}\\
  \label{eq:Hodge-duals}
\end{equation}
Applying this formula to computing the Hodge dual of $0$-forms, it follows in
particular that $\star1=\bfs{\epsilon}$.

An important property of the Hodge dual is that for any $p$-form
\begin{equation}
  \star\star\bfs{A} = (-1)^{1+p(4-p)}\bfs{A}\,,
\end{equation}
which implies
\begin{equation}
  \star^{-1}\bfs{A} = (-1)^{1+p(4-p)}\star\bfs{A}\,.
\end{equation}

Another property of $p$-forms which is fundamental for the present work is that
they are natural integrands over $p$-dimensional (sub-)manifolds of
$\mathcal{M}$. In particular, if a $p$-dimensional submanifold of $\mathcal{M}$
is further divided into a set of non-overlapping $p$-dimensional regions, a
$p$-form $\bfs{A}$ naturally establishes a map from this set to the set of real
numbers. If $\mathcal{S}_1$ and $\mathcal{S}_2$ are such regions, then
\begin{equation}
  \bfs{A}[\mathcal{S}_1]=\int_{\mathcal{S}_1}\bfs{A}
\end{equation}
and
\begin{equation}
  \bfs{A}[\mathcal{S}_1\cup\mathcal{S}_2]=
  \int_{\mathcal{S}_1}\bfs{A}+\int_{\mathcal{S}_2}\bfs{A}\,.
\end{equation}
Note in particular that the integral of $\bfs{\epsilon}$ over a portion of
$\mathcal{M}$ is nothing but the volume of that portion, hence the name volume
form for $\bfs{\epsilon}$.

We can then state the modern version of Stokes' theorem, which generalizes the
well known theorems of vector calculus by Green, Stokes and Gauss. It allows to
relate integrals over a general submanifold $\mathcal{S}$ of $\mathcal{M}$ to
integrals over its boundary $\partial\mathcal{S}$
\begin{equation}
  \int_{\mathcal{S}}\ext\bfs{A}=\int_{\partial\mathcal{S}}\bfs{A}\,.
  \label{eq:stokes}
\end{equation}
\eqref{eq:stokes} too has a fundamental importance for this work.

Finally, it can be useful to restate standard vector-calculus operators in terms
of differential forms and exterior calculus operators, \eg:
\begin{align}
  \textnormal{grad}(f) &= \overrightarrow{\ext f}\,,\\
  \label{eq:ddiv}
  \textnormal{div}(\bfs{u}) &= -\star^{-1}\ext\star\widetilde{\bfs{u}}\,,\\
  \label{eq:dcurl}
  \textnormal{curl}(\bfs{u}) &= \overrightarrow{\star\ext\widetilde{\bfs{u}}}\,.
\end{align}
In these expressions $f$ is a generic scalar field (or equivalently a $0$-form),
and $\bfs{u}$ a generic vector; an arrow is used to denote the operation of
transforming a differential $1$-form to its dual vector, and a tilde to denote
the inverse operation.

\section{Hodge dual of the Nester-Witten form in terms of the connection}
\label{sec:omega_to_dualU}
In order to obtain equation \eqref{eq:dual_u_from_omega},
we start from the definition of the Nester-Witten form
\eqref{eq:nester-witten-definition}, which can also be written as
\begin{equation}
	\bfs{u}_a = -\frac{1}{2}\omega^{bc}_{\ \ d} \base^{d}
	\wedge\bfs{\Sigma}_{abc}\,.
\end{equation}
Using the identity \eqref{eq:base_wedge_Sigma}, we obtain
\begin{equation}
\bfs{u}_a = -\frac{3}{2}\omega^{bc}_{\ \ d}
 \delta^{d}_{\ [a} \bfs{\Sigma}_{bc]} \,.
\end{equation}
Expanding the antisymmetric brackets,
\begin{align}
\begin{split}
\bfs{u}_a = -\frac{1}{4} \left[ \right.
& \omega^{bc}_{\ \ a} \bfs{\Sigma}_{bc}
 +\omega^{bd}_{\ \ d} \bfs{\Sigma}_{ab}
 +\omega^{dc}_{\ \ d} \bfs{\Sigma}_{ca}\\
-&\omega^{bc}_{\ \ a} \bfs{\Sigma}_{cb}
 -\omega^{dc}_{\ \ d} \bfs{\Sigma}_{ac}
 -\omega^{bd}_{\ \ d} \bfs{\Sigma}_{ba}
\left. \right] \,,
\end{split}
\end{align}
and renaming indices to factor out $\bfs{\Sigma}_{bc}$,
\begin{equation}
\bfs{u}_a = -\frac{1}{2}
\left(\omega^{[bc]}_{\ \ \ a}
    +\delta^{b}_{\ a}\omega^{[cd]}_{\ \ \ d}
    -\delta^{c}_{\ a}\omega^{[bd]}_{\ \ \ d}
     \right)\bfs{\Sigma}_{bc} \,.
\end{equation}
From the definition of the hypersurface forms \eqref{eq:definition_sigma}
and the formula to obtain the components of the Hodge dual
\eqref{eq:Hodge-duals}, it follows that the expression in
parenthesis equals the components of $\star\bfs{u}_a$,
as stated in equation \eqref{eq:dual_u_from_omega}.

\section{Derivation of evolution equation for $\sqrt{\gamma}$}
\label{sec:dt_sqrtgamma_derivation}
As mentioned in Section \ref{sec:closing_system}, one can obtain
an evolution equation for $\sqrt{\gamma}$ in a conservative form
by taking the exterior derivative of the hypersurface form $\bfs{\Sigma}_{\hat{0}}$
orthogonal to $-n_\mu$. This form is identical to the Hodge dual of $\base^{\hat{0}}$,
and in a coordinate basis it has components
\begin{equation}
(\bfs{\Sigma_{\hat{0}}})_{\mu\nu\lambda} = -\varepsilon_{\alpha\mu\nu\lambda} n^\alpha \,.
\end{equation}
Similarly as in Section \ref{sec:ext_systems},
the components of its exterior derivative will take the form of a conservation
equation for $n^\alpha$,
\begin{equation}
\left(\frac{1}{\sqrt{-g}}\partial_\mu \sqrt{-g} n^\mu \right)\bfs{\Sigma}
= \ext \bfs{\Sigma_{\hat{0}}}
\end{equation}
or taking a single component (for a 4-form, all of them are equal up
to a factor $-1$),
\begin{equation}
\label{eq:sqtrgamma_evol_lhs}
\partial_t \sqrt{\gamma}-\partial_i \sqrt{\gamma}\beta^i
= (\ext \bfs{\Sigma_{\hat{0}}})_{0123}
\end{equation}
To find an expression for the right hand side, we recall the definition
of hypersurface forms \eqref{eq:definition_sigma} and write
\begin{align}
\begin{split}
\bfs{\Sigma}_{\hat{0}} =& \varepsilon_{\hat{0}\hat{\mu}\hat{\nu}\hat{\lambda}}
\base^{\hat{\nu}}\wedge\base^{\hat{\mu}}\wedge\base^{\hat{\lambda}} \\
=& -\varepsilon_{\hat{i}\hat{j}\hat{k}}
\base^{\hat{i}}\wedge\base^{\hat{j}}\wedge\base^{\hat{k}} \,.
\end{split}
\end{align}
Now we take the exterior derivative and using the Leibniz rule for the
exterior product \eqref{eq:leibniz_ext}
and the definition of $\bfs{F}^{\hat{a}}$ \eqref{eq:F_definition} to obtain
\begin{equation}
\ext \bfs{\Sigma}_{\hat{0}} = -\varepsilon_{\hat{i}\hat{j}\hat{k}} 
\bfs{F}^{\hat{i}}\wedge\base^{\hat{j}}\wedge\base^{\hat{k}} \,.
\end{equation}
For the last step, we need to obtain the component  $0123$
of this form in a coordinate frame.
This reads
\begin{align}
\begin{split}
(\ext \bfs{\Sigma}_{\hat{0}})_{0123} =& -\frac{1}{24}\varepsilon_{\hat{i}\hat{j}\hat{k}}
F^{\hat{i}}_{\ \hat{\mu}\hat{\nu}}
\delta^{\hat{\mu}\hat{\nu}\hat{j}\hat{k}}_{\hat{\alpha}\hat{\beta}\hat{\gamma}\hat{\delta}} A^{\hat{\alpha}}_{\ 0} A^{\hat{\beta}}_{\ 1} A^{\hat{\gamma}}_{\ 2} A^{\hat{\delta}}_{\ 3} \\
=&-\frac{1}{24}\varepsilon_{\hat{i}\hat{j}\hat{k}}
F^{\hat{i}}_{\ \hat{\mu}\hat{\nu}}
\delta^{\hat{\mu}\hat{\nu}\hat{j}\hat{k}}_{\hat{0}\hat{l}\hat{m}\hat{n}} A^{\hat{0}}_{\ 0} A^{\hat{l}}_{\ 1} A^{\hat{m}}_{\ 2} A^{\hat{n}}_{\ 3}
\end{split}
\end{align}
where the renaming of indices follows from the fact that $A^{\hat{0}}_{\ i}=0$,
so the lower indices of the $\delta$ will be a permutation of the upper ones
only if the remaining indices are different from zero.
Since the hatted latin indices cannot take the value $\hat{0}$, many components
of the delta are eliminated, and, taking into account the antisymmetry on the last
two indices of $F^{\hat{i}}_{\ \hat{\mu}\hat{\nu}}$, we arrive at
\begin{equation}
\begin{split}
(\ext \bfs{\Sigma}_{\hat{0}})_{0123} =&-\frac{1}{12}\varepsilon_{\hat{i}\hat{j}\hat{k}}
F^{\hat{i}}_{\ 0\hat{p}}
\delta^{\hat{p}\hat{j}\hat{k}}_{\hat{l}\hat{m}\hat{n}}
A^{\hat{l}}_{\ 1} A^{\hat{m}}_{\ 2} A^{\hat{n}}_{\ 3} \\
=&-\frac{1}{2}\varepsilon_{\hat{i}\hat{j}\hat{k}}
F^{\hat{p}}_{\ 0\hat{p}}
A^{\hat{i}}_{\ 1} A^{\hat{j}}_{\ 2} A^{\hat{k}}_{\ 3} \\
=&-\frac{1}{2}\sqrt{\gamma}
F^{\hat{p}}_{\ 0\hat{p}}
\end{split}
\end{equation}
where the second step follows from the fact that $\hat{p}\hat{j}\hat{k}$
can be a permutation of $\hat{i}\hat{j}\hat{k}$ only if $\hat{p}=\hat{i}$
and the third from the definition of the determinant.
Now we can evaluate the sum by contracting \eqref{eq:F_components}
and then using the constitutive relation \eqref{eq:Ek_DH},
\begin{equation}
    \label{eq:sqtrgamma_evol_rhs}
	(\ext \bfs{\Sigma}_{\hat{0}})_{0123}
	          = -\frac{1}{2}\sqrt{\gamma} E^{\hat{i}}_{\ \hat{i}}
	          = \frac{5}{2} D_{\hat{k}}^{\ \hat{k}} \,.
\end{equation}
Finally, equating \eqref{eq:sqtrgamma_evol_lhs} and
\eqref{eq:sqtrgamma_evol_rhs}, we obtain equation \eqref{eq:sqrtgammaevol}.

\end{document}